\documentclass[12pt]{iopart}
\usepackage{iopams}  
\usepackage{graphicx}
\usepackage{hyperref}

\expandafter\let\csname equation*\endcsname\relax
\expandafter\let\csname endequation*\endcsname\relax

\usepackage{amsmath}
\usepackage{amssymb}

\begin{document}

\title[Influence of committed volunteers on helping behavior in emergency evacuations]{Influence of committed volunteers on helping behavior in emergency evacuations}

\author{Jaeyoung Kwak$^{1}$, Michael H. Lees$^2$, Wentong Cai$^1$, A. Reza Pourghaderi$^{3,5}$, and Marcus E.H. Ong$^{4,5}$}
\address{$^1$ School of Computer Science and Engineering, Nanyang Technological University, Singapore}
\address{$^2$ Informatics Institute, University of Amsterdam, The Netherlands}
\address{$^3$ Health Systems Research Center (HSRC), Singapore Health Services, Singapore}
\address{$^4$ Department of Emergency Medicine, Singapore General Hospital, Singapore}
\address{$^5$ Health Services and Systems Research (HSSR), Duke-NUS Medical School, Singapore}

\vspace{10pt}
\begin{indented}
\item[]Date: \today
\end{indented}

\begin{abstract} 
We study how the presence of committed volunteers influences the collective helping behavior in emergency evacuation scenarios. In this study, committed volunteers do not change their decision to help injured persons, implying that other evacuees may adapt their helping behavior through strategic interactions. An evolutionary game theoretic model is developed which is then coupled to a pedestrian movement model to examine the collective helping behavior in evacuations. By systematically controlling the number of committed volunteers and payoff parameters, we have characterized and summarized various collective helping behaviors in phase diagrams. From our numerical simulations, we observe that the existence of committed volunteers can promote cooperation but adding additional committed volunteers is effective only above a minimum number of committed volunteers. This study also highlights that the evolution of collective helping behavior is strongly affected by the evacuation process.
\end{abstract}

%
\vspace{2pc}
\section{Introduction}
\label{intro}
Pedestrian emergency evacuation has generated considerable research interest for the past several years. A clear understating of emergency evacuation process is vital for evacuating people to the place of safety in emergency situations such as fire and hazardous chemical spills. Various phenomena relevant to emergency evacuations have been studied, for instance faster-is-slower effect~\cite{Helbing_Nature2000} and crowd turbulence~\cite{Helbing_PRE2007}. Those phenomena have studied based on different approaches including empirical observations~\cite{Helbing_PRE2007, Helbing_EPJDataSci2012}, controlled experiments~\cite{Kelley_1965, Pastor_PRE2015}, and simulations~\cite{Helbing_Nature2000, Yu_PRE2007, Sticco_PRE2017}. In the majority of studies, it has been reported that people tend to move faster than their normal speed and behave in an individualistic manner during emergency evacuations, potentially increasing the risk of stampede~\cite{Helbing_Nature2000, Helbing_EPJDataSci2012, Kelley_1965}. In view of this, various pedestrian emergency evacuation studies were performed, for instance, predicting total evacuation time in a class room~\cite{Guo_TrB2012} and preparing an optimal evacuation plan for a large scale pedestrian facility~\cite{Abdelghany_EJOR2014}. 

However, other studies have presented evidence that evacuees may help injured persons to escape the place of danger, for instance, the WHO concert disaster occurred on December 3, 1979 in Cincinnati, Ohio, United States~\cite{Johnson_1987} and the 2005 London bombings in the United Kingdom~\cite{Drury_2009}. There is existing research that investigated helping behavior in emergency evacuation by means of pedestrian simulations. For instance, von Sivers~\textit{et~al.} \cite{vonSivers_PED2014,vonSivers_SafetySci2016} applied social identity theory to pedestrian simulation in order to simulate helping behavior observed in the 2005 London bombings. In their studies, they assumed that all the evacuees share the same social identify which makes them willing to help others rather than be selfish.

Although some studies have studied helping behavior in emergency evacuations through simulation, little attention has been paid to the strategic interactions among evacuees. When an evacuee is helping an injured person to escape, the evacuee's helping behavior can be seen as an attempt to increase a collective good. This is especially true when there are not enough dedicated rescue personnel, in such circumstances evacuees may help one another. At the same time, helping an injured person can be a costly activity because the volunteering evacuee spends extra time and increases their personal risk. If an evacuee feels that helping behavior is a costly behavior for them, they might not help the injured person. 

Game theory is a commonly used approach to predict how individuals decide their strategy in response to others' strategy. Under game theoretic assumptions, individuals are likely to select a strategy in a way to maximize their own payoff. Game theoretical models have provided insights how human cooperation emerges from the interactions among such selfish individuals~\cite{Nowak_Science2006, Perc_PhysRep2017}. In addition, game theoretical models have demonstrated potential in strategy development for addressing social problems including criminal activity detection~\cite{Perc_PLOS2013, Helbing_JStatPhys2015} and vaccination~\cite{Helbing_JStatPhys2015, Bauch_PNAS2004, Liu_PRE2012, Chapman_PsySci2012}. Various game theoretical models have been also applied for emergency evacuation simulations, for example, the prisoner's dilemma game~\cite{Bouzat_PRE2014}, the snowdrift game~\cite{Shi_PRE2013}, the spatial game~\cite{Heliovaara_PRE2013,Mitsopoulou_JCS2019}, and the evolutionary game~\cite{Hao_PRE2011}. 

In our previous work~\cite{Kwak_ICCS2020}, we employed the volunteer's dilemma game~\cite{Diekmann_JCR1985,Diekmann_SF2016} to study the impact of volunteering cost on collective helping behavior in emergency evacuations. In Ref.~\cite{Kwak_ICCS2020}, we observed different patterns of collective helping behaviors when varying the volunteering cost parameter. Nevertheless, little is known about how the existence of committed volunteers can lead to the emergence of collective helping behavior in emergency evacuations. The committed volunteers are insensitive to the volunteering cost, so they do not change their decision to help the injured persons. One can imagine that, through strategic interactions, committed volunteers may influence other evacuees to also help injured people, thereby facilitating the spread of cooperation in the population.

In the area of game theoretic modeling, a considerable number of studies have examined the influence of committed minorities who are immune to strategy updates. For instance, Nakajima and Masuda~\cite{Nakajima_JMathBio2015} studied the influence of committed minorities in finite populations on the fixation time for two-strategy matrix games. In another study, Liu~\textit{et~al.}~\cite{Liu_PRE2012} examined how the committed vaccinators promote vaccine uptake for well-mixed and spatially structured populations. Recently, Cardillo and Masuda~\cite{Cardillo_PRR2020} demonstrated the critical mass effect due to the existence of committed minorities. While those studies reported the enhancement of cooperation by committed minorities, they do not account for change in population in their studies. It has been pointed out that time-varying population size in evolutionary games might display collective dynamics in a qualitatively different way from that predicted by fixed, finite population models~\cite{Melbinger_PRL2010, Cremer_PRE2011, McAvoy_TPB2015}. Other studies also suggest that interactions between evolutionary game dynamics and network growth might affect the promotion of cooperation~\cite{Poncela_PLOS2008, Poncela_NJP2009, Perc_BioSys2010}.  

In this work, we perform numerical simulations for a room evacuation scenario in which two volunteers are required to move an injured person to the place of safety. In our simulations, evacuees update their strategy after strategic interactions with other evacuees within their sensory range while the evacuees move to the place of safety. Hence, the strategic interactions between evacuees and the change of their position driven by the evacuation process are coupled together in our study. We develop an evolutionary game theoretic model to study the collective helping behavior in emergency evacuations emerging from the strategic interactions among evacuees with the existence of committed volunteers. The evolutionary game theoretic model and its numerical simulation setups are described in Section~\ref{section:model}. We characterize the numerical simulation results depending on the number of committed volunteers and the payoff parameters in Section~\ref{section:results}. In addition, we provide an explanation for the general tendency observed from the simulation results by looking into the fractions of neighbors susceptible to strategy adaptation. Finally, we summarize our findings and present concluding remarks in Section~\ref{section:conclusion}.

\section{Model}
\label{section:model}
Our agent-based model consists of a game theoretical model and a movement model. The game theoretical model evaluates the probability that a bystander would turn into a volunteer helping an injured person. The evolutionary game model enables us to reflect the behavioral changes of an individual based on the behavior of other evacuees. The movement model calculates the sequence of pedestrian positions for each simulation time step. In this section, we explain our game theoretical model, pedestrian movement model, and the numerical simulation setup.

\subsection{Evolutionary game model}
\label{section:EvolGame}
\begin{figure}[!t]
	\centering
	\begin{indented}
		\item[]
		\includegraphics[width=6cm]{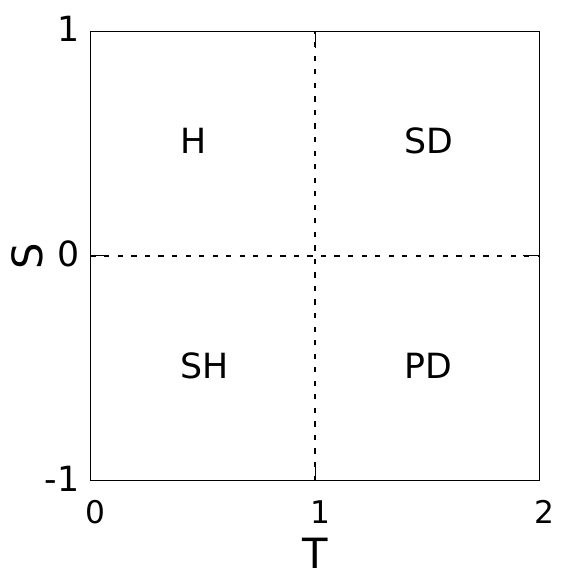}
	\end{indented}
	\caption{Schematic representation of four different games on $(T, S)$ space: the upper left quadrant corresponds to the harmony (H) game ($1 > T > 0$ and $1 > S > 0$), the upper right one to the snowdrift (SD) game ($2 > T > 1$ and $1 > S > 0$), the lower left one to the stag hunt (SH) game ($1 > T > 0$ and $0 > S > -1$), and the lower right one to the prisoner's dilemma (PD) game ($2 > T > 1$ and $0 > S > -1$).}
	\label{fig:Schematic_T-S} 
\end{figure}

We employ an evolutionary game model to study the behavioral change of players influenced by other players. As in our previous study~\cite{Kwak_ICCS2020}, we consider two strategies of ambulant pedestrians: volunteer (C) and bystander (D). A volunteer (C) helps an injured person to evacuate whereas a bystander (D) does not help the injured person. In line with the two-player two-strategy games, the payoff of individual $i$'s strategy can be presented in a payoff matrix in Table~\ref{table:payoff_2player}. When individual $i$ meets individual $j$, there are four different payoffs depending on the strategy of each individual. If both the individuals are volunteers (C), they receive a reward for mutual cooperation, denoted by $R$. If both are bystanders (D), their payoff is $P$ which is punishment for mutual defection. When bystander $i$ is a volunteer (C) and bystander $j$ is a bystander (D), bystander $i$ receives the sucker's payoff $S$ which is associated with the unreciprocated cooperation cost~\cite{Smaldino_AmNat2013}. At the same time, bystander $j$ receives $T$ which reflects the temptation to defect. In line with previous work~\cite{Cardillo_PRR2020, Santos_PNAS2006, Helbing_PLOS2010, Szabo_JTB2012, Szabo_PhysRep2016}, we employ the rescaled payoff setup by which we normalize the mutual cooperation payoff to 1 and mutual defection to 0. By doing so, we can characterize the games by exploring the $(T, S)$ parameter space in the range of $0 \leq T \leq2$ and $-1 \leq S \leq 1$, see Figure~\ref{fig:Schematic_T-S}. 

As shown in Figure~\ref{fig:Schematic_T-S}, the $(T, S)$ space can be divided into four areas: the harmony (H) game ($1 > T > 0$ and $1 > S > 0$), the snowdrift (SD) game ($2 > T > 1$ and $1 > S > 0$), the stag hunt (SH) game ($1 > T > 0$ and $0 > S > -1$), and the prisoner's dilemma (PD) game ($2 > T > 1$ and $0 > S > -1$). Dividing the $(T, S)$ space into different areas is useful to interpret different forms of strategic interaction. Although mutual cooperation is preferred over the unilateral cooperation for all games, i.e., $R > S$, the preferred strategy for players is different in different games~\cite{Santos_PNAS2006, Macy_PNAS2002, Traulsen_PRE2006a}. For instance, players are likely to select unilateral defection rather than mutual cooperation (i.e., $T > R$) in the SD game, while mutual defection is preferred over unilateral cooperation (i.e., $P > S$) in the SH game. In the PD game, defection is the most tempting strategy for every player (i.e., $T > R$ and $P > S$) and as such the game ends up in the situation in which all players select mutual defection. 

\begin{table}
	\caption{\label{table:payoff_2player}Payoff of a volunteer (C) and a bystander (D) in a two-player game. Here, $R$ is the reward for mutual cooperation, $P$ is the punishment for mutual defection, $T$ is the temptation to defect, and $S$ is the sucker's payoff. The entities indicate the payoff for the row player.\\}
	\begin{indented}
		\lineup
		\item[]
		\begin{tabular}{ccc}
		\br
		& Volunteer & Bystander\\
		& (C) & (D)\\
		\mr
		Volunteer (C) & $R$ & $S$\\
		Bystander (D) & $T$ & $P$\\
		\br
		\end{tabular}
	\end{indented}
\end{table}

We assume that two volunteers are required to move an injured person from the room to the place of safety. A lonely volunteer cannot move the injured person by himself/herself, so the lonely volunteer will try to find another volunteer who will move the injured person together. A bystander near the lonely volunteer might decide to become a volunteer to help the lonely volunteer to move the injured person together. In contrast, the lonely volunteer might give up finding another volunteer and turn into a bystander. 

In each round, ambulant pedestrians collect payoffs from their neighbors within their sensory range and assign the average of the collected payoffs from each to be their individual payoff~\cite{Cardillo_PRR2020, Traulsen_JTB2007a, Kleineberg_NatureComm2017}. Formally, the payoff of individual $i$ at time $t$, $u_i(t)$ is given as by
\begin{equation}\label{eq:payoff_i}
	u_i(t) = \frac{1}{\left| M_{i}(t) \right|} \sum_{j \in M_{i}(t)}^{} a_{ij},
\end{equation}
where $M_{i}(t)$ is the set of focal individual $i$'s neighbors at time $t$ within the sensory range. The awarded payoff from the interaction with individual $j$ is denoted by $a_{ij}$, which is given by the payoff matrix in Table~\ref{table:payoff_2player}. After evaluating individuals' payoff, ambulant pedestrian $i$ randomly selects a neighbor $j$ and adopts neighbor $j$'s strategy according to the strategy adaptation probability~\cite{Traulsen_PRE2006a, Blume_GEB1993, Szabo_PRE1998, Mobilia_PRE2012}:
\begin{equation}\label{eq:prob_adopt}
	p(u_{ij}) = \frac{1}{1+\exp (\beta u_{ij})},
\end{equation}
\noindent
where $\beta$ reflects selection intensity in the game dynamics and $u_{ij} = u_i-u_j$ is payoff difference. If $\beta = 0$, the probability $p(u_{ij})$ becomes 0.5, indicating completely random decision making. In case of $\beta \rightarrow \infty$, the probability $p(u_{ij})$ becomes $1$ if $u_i < u_j$, and $p(u_{ij})$ is $0$ when $u_i > u_j$.

\subsection{Social force model}
\label{section:SFM}

We describe the movement of pedestrians based on the social force model~\cite{Helbing_PRE1995}. The position and velocity of each pedestrian $i$ at time $t$, denoted by $\vec{x}_i(t)$ and $\vec{v}_i(t)$, are updated according to the following equations:
\begin{equation}\label{eq:SFM_v}
	\frac{\mathrm{d} \vec{x}_i(t)}{\mathrm{d} t} = \vec{v}_i(t)
\end{equation}
\noindent 
and
\begin{equation}\label{eq:SFM_acc}
	\frac{\mathrm{d} \vec{v}_i(t)}{\mathrm{d} t} = \vec{f}_{i, d} + \sum_{j\neq i}^{ }{\vec{f}_{ij}} + \sum_{B}^{ }{\vec{f}_{iB}}.
\end{equation}
In Eq.~(\ref{eq:SFM_acc}), the driving force term $\vec{f}_{i,d}$ describes the tendency of pedestrian $i$ to move toward their destination. The repulsive force terms $\vec{f}_{ij}$ and $\vec{f}_{iB}$ reflect the tendency to maintain a certain distance from other pedestrian $j$ and the boundary $B$, e.g., walls and obstacles. We refer the readers to \ref{section:SFM_details} for a more detailed description of the presented social force model. 

\subsection{Numerical simulation setup}
\label{section:setup}

\begin{figure}[!t]
	\centering
	\begin{indented}
		\item[]
		\includegraphics[width=8.5cm]{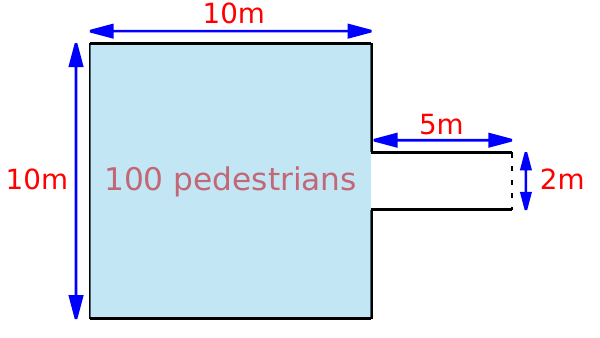}		
	\end{indented}
	\caption{Schematic depiction of the emergency evacuation simulation setup. 100 pedestrians are placed in a 10m$\times$10m room indicated by a blue shade area. Pedestrians are leaving the room through an exit corridor which is 5~m long and 2~m wide. The place of safety is set on the right, outside of the exit corridor.}
	\label{fig:room_layout}  
\end{figure}

Each pedestrian is modeled as a circle with radius $r_i = 0.2$~m. We placed $N_0 = 100$ pedestrians in a 10m$\times$10m room indicated by the blue shaded area in Figure~\ref{fig:room_layout}. Pedestrians leave the room through an exit corridor which is 5~m long and 2~m wide. The place of safety is set on the right, beyond the exit corridor. The pedestrians' movement and position are updated according to the social force model defined in Eq.~(\ref{eq:SFM_acc}).

There are $N_i=10$ injured persons who need help in escaping the room and $N=N_0-N_i=90$ ambulant pedestrians who are either volunteers or bystanders. The number of volunteers at time $t$, $N_{C}(t)$, is given as a sum of the number of committed volunteers $N_{C, Z}$ and uncommitted volunteers $N_{C, S}(t)$, i.e., $N_{C}(t) = N_{C, Z}+N_{C, S}(t)$. While the number of committed volunteers $N_{C, Z}$ is constant during the simulation, the uncommitted volunteers can change their strategy, thus the number of uncommitted volunteers $N_{C, S}(t)$ is changing over the time. The number of defectors is indicated as $N_{D}(t) = N-N_{C}(t)$. At the start of each evacuation simulation, one volunteer is selected for each injured person within the injured person's sensory range $l_s$, accordingly there are $N_{C}(0) = 10$ initial volunteers. Among those initial volunteers, we then select a certain number of committed volunteers, consequently $N_{C, Z} \leq N_{C}(0)$. As described in Section~\ref{section:EvolGame}, we assume that an injured person needs two volunteers, thus the maximum number of volunteers $N_{C,\max}$ is 20. We perform simulations for different numbers of committed volunteers $N_{C, Z}$. The number of volunteers $N_{C}(t)$ evolves according to the result of strategic interaction with other evacuees. 

\begin{table}[!t]
	\centering
	\caption{\label{table:EvolGame_parameter}Model parameters\\}
	\begin{indented}
		\lineup
		\item[]
		\begin{tabular}{@{}lll}
			\br
			Model parameter & symbol & value \\
			\mr
			mutual cooperation payoff	& $R$		& 1\\
			mutual defection payoff		& $P$		& 0\\
			temptation payoff			& $T$		& [0, 2]\\
			sucker's payoff				& $S$		& [-1, 1]\\
			selection intensity			& $\beta$	& 10\\	
			sensory range				& $l_s$		& 3\\	
			the number of initial volunteers & $N_{C}(0)$ & 10\\			
			the number of committed volunteers & $N_{C, Z}$ & [0, 10]\\	
			the maximum number of volunteers & $N_{C,\max}$ & 20\\				
			\br
		\end{tabular}
	\end{indented}
\end{table}

All the ambulant pedestrians play the presented evolutionary game every $0.2$ seconds. Similar to pedestrian goal selection and path navigation behaviors, we assume that the evolutionary game is a macroscopic behavior so the players update their strategies several times per second~\cite{Heliovaara_PRE2013, Zhong_AAMAS2016}. Each ambulant pedestrian $i$ randomly selects pedestrian $j$ within its sensory range $l_s$ and then evaluates the probability of switching to pedestrian $j$'s strategy according to Eq.~(\ref{eq:prob_adopt}). Once a bystander interacts with a lonely volunteer and decides to cooperate with the lonely volunteer to rescue an injured person, the bystander turns into a volunteer. The new volunteer shifts their desired walking direction toward the position of injured person. After arriving at the injured person, the volunteer will evacuate the injured person with the peer volunteer after a preparation time of 60~s. On the other hand, if a lonely volunteer decides to become a bystander, they change the desired walking direction vector toward the exit. To examine the influence of committed volunteers, we control the number of committed volunteers $N_{C,Z}$ along with the game payoff parameters $T$ and $S$. By controlling the value of $T$ and $S$, we can study different setups of strategic interactions. According to previous studies~\cite{Kwak_ICCS2020, Lin_PRE2018}, we choose the value of sensory range $l_s = 3$~m. The selection intensity $\beta$ is set as 10 based on previous work~\cite{Szabo_JTB2012, Amaral_PRE2018, Wang_PRE2010}.

\section{Results and discussions}
\label{section:results}

\subsection{General tendency}
\label{section:Results_General}

In this section, we first perform an analysis of the stationary state value of cooperation level for the case of fixed population size. Next we compare the analytical results with that from the numerical simulation to quantify the impact of evacuation process on the collective helping behavior.

Based on previous studies~\cite{Nakajima_JMathBio2015, Cardillo_PRR2020, Masuda_SciRep2012}, we consider a well-mixed population model with a fixed, finite population consisting of $N$ ambulant pedestrians. That is, the number of pedestrians is constant over the time, indicating that the decline in population due to evacuation process is not considered. The evolution of the number of cooperators, $dN_{C}(t)/dt$, can be described in terms of the rate of increase $\Gamma_{+}$ and the rate of decrease $\Gamma_{-}$, 
\begin{equation}
	\label{eq:Nc_dot} 
	\frac{dN_{C}(t)}{dt} = \Gamma_{+}-\Gamma_{-}.	
\end{equation}
\noindent
The increase rate $\Gamma_{+}$ and the decrease rate $\Gamma_{-}$ are given as 
\begin{equation}
	\label{eq:Gamma_plus} 
	\Gamma_{+} = \frac{(N_{C,Z}+N_{C,S})N_{D}}{N(N-1)} \frac{1}{1+\exp(\beta u_{DC})},
\end{equation}
\noindent
and 
\begin{equation}
	\label{eq:Gamma_minus} 
	\Gamma_{-} = \frac{N_{C,S}N_{D}}{N(N-1)} \frac{1}{1+\exp(\beta u_{CD})}.
\end{equation}
\noindent
Here, $N_{C,S}$ is the number of uncommitted volunteers and $N_{D}$ is the number of defectors. The payoff differences $u_{DC} = u_D-u_C$ and $u_{CD} = u_C-u_D$ denote the probability of switching strategy from $D$ to $C$ and $C$ to $D$, respectively. In Eq.~(\ref{eq:Gamma_plus}), $(N_{C,Z}+N_{C,S})N_{D}/N(N-1)$ indicates the pair selection probability by which a pair of cooperator and defector is selected. The pair selection probability in Eq.~(\ref{eq:Gamma_minus}) is given as $N_{C,S}N_{D}/N(N-1)$, reflecting that the committed volunteers are not involved in the decrease of cooperators. The payoffs associated with cooperators $u_C$ and defectors $u_D$ are given as 
\begin{equation}
	\label{eq:Payoff_C}
		u_C = \frac{(N_{C,Z}+N_{C,S})-1}{N-1}R + \frac{N_{D}}{N-1}S		
\end{equation}
\noindent
and
\begin{equation}
	\label{eq:Payoff_D}
		u_D = \frac{(N_{C,Z}+N_{C,S})}{N-1}T + \frac{N_{D}-1}{N-1}P.	
\end{equation}
\noindent
Note that the committed volunteers can induce defectors to become cooperators, but do not change their strategy to defection.   

Based on Eqs.~(\ref{eq:Gamma_plus})~and~(\ref{eq:Gamma_minus}), we rewrite Eq.~(\ref{eq:Nc_dot}) as  
\begin{equation}
	\label{eq:Nc_dot_detail} 
	\begin{split}
		\frac{dN_{C}(t)}{dt}
		&= \frac{(N_{C,Z}+N_{C,S})N_{D}}{N(N-1)}\frac{1}{1+\exp(\beta u_{DC})}-\frac{N_{C, S}N_{D}}{N(N-1)}\frac{1}{1+\exp(\beta u_{CD})}\\
		&= \frac{N_{D}}{N(N-1)} \frac{(N_{C,Z}+N_{C,S})- N_{C,S}\exp(\beta u_{DC})}{1+\exp(\beta u_{DC})}.
	\end{split}
\end{equation}    

To quantify the level of collective helping behavior, based on Ref.~\cite{Bouzat_PRE2014}, we define the cooperation level $\rho_{c}$~$\in$~$[-1, 1]$ as
\begin{equation}\label{eq:fraction_cooperation}
	\rho_{c} = \frac{N_{C}-N_{C}(0)}{N_{C}(0)}.
\end{equation}
\noindent
A positive value of $\rho_{c}$ suggests that the final number of volunteers $N_{C}$ is greater than the initial one, i.e., $N_{C}(0)$, inferring that some lonely volunteers successfully found peer volunteers. In case of $\rho_{c} = 1$, all the lonely volunteers find a peer and hence all the injured persons are rescued. A negative value indicates that some lonely volunteers changed their mind to become bystanders. In case of $\rho_{c} = -1$, all the lonely volunteers turn into bystanders, resulting in none of the injured persons being rescued. 

\begin{figure*}[!t]
	\centering
	\begin{tabular}{c}
		\includegraphics[width=15.5cm]{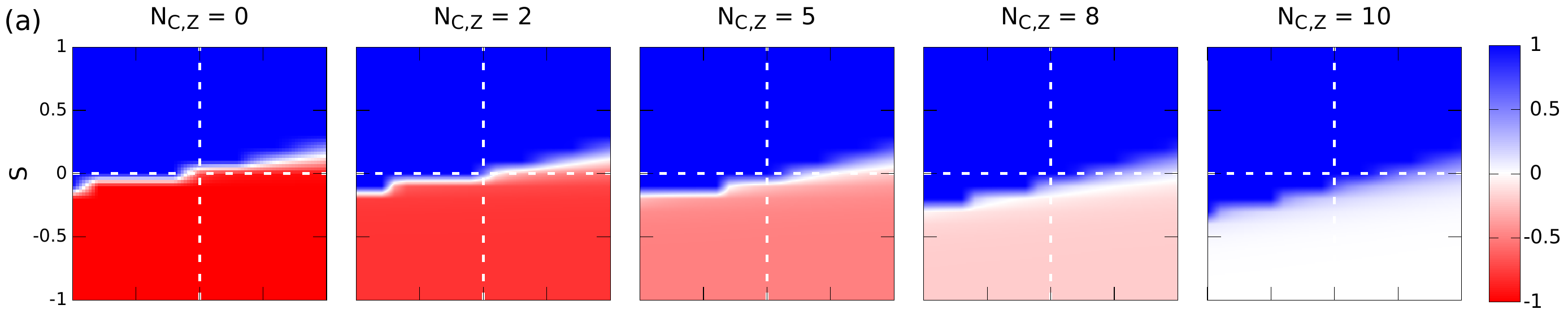}\vspace{-0.1cm}\\
		\includegraphics[width=15.5cm]{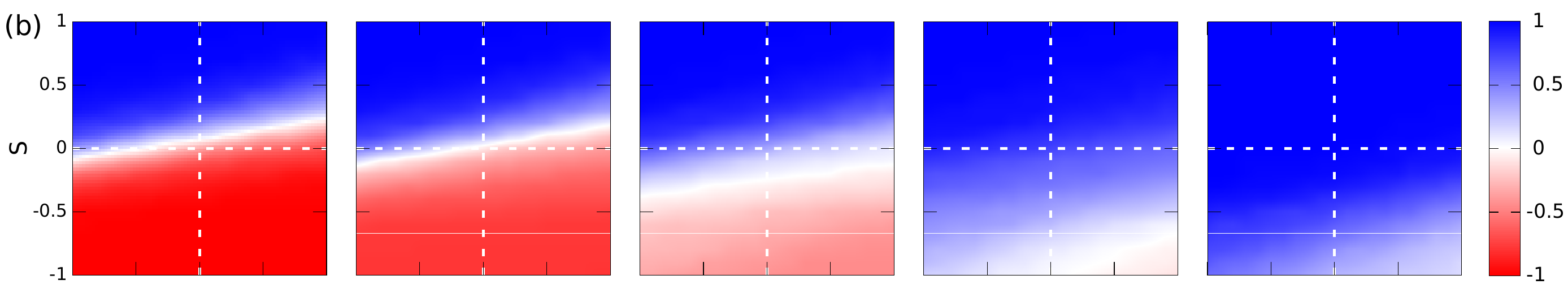}\vspace{-0.1cm}\\
		\includegraphics[width=15.5cm]{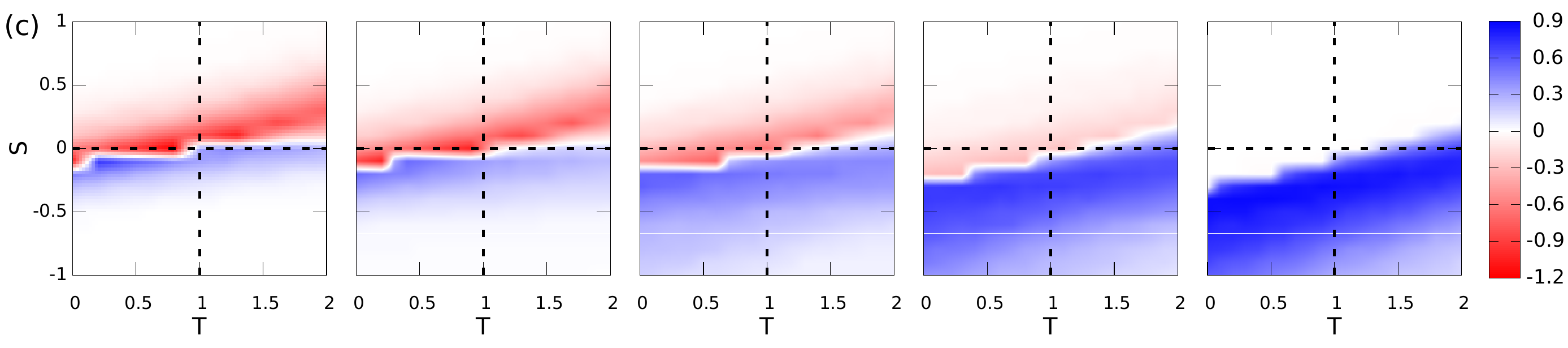}\vspace{-0.3cm}\\
	\end{tabular}	
	\caption{The average stationary state value of cooperation level estimated for (a) the fixed population case (i.e., $\rho_{c, \textnormal{fixed}}$) and (b) the emergency evacuation case (i.e., $\rho_{c, \textnormal{evac}}$). In (a) and (b), the cooperation level $\rho_{c}$ is 1 (blue) if all the lonely volunteers eventually find their peers, suggesting that all the injured persons are rescued. In contrast, $\rho_{c} = -1$ (red) indicates that all the lonely volunteers turn into bystanders, indicating that none of the injured persons are rescued. The difference in cooperation level  between the fixed population size case and emergency evacuation case (i.e., $\Delta \rho_{c} = \rho_{c, \textnormal{evac}}-\rho_{c, \textnormal{fixed}}$) is shown in (c). A negative $\Delta \rho_{c}$ (red) suggests that the evacuation process has a negative impact on the cooperation level, while a positive $\Delta \rho_{c}$ (blue) implies that the evacuation process contributes to further increase the cooperation level. Each panel corresponds to different numbers of committed volunteers $N_{C, Z}$. Results are generated with parameter values listed in Table~\ref{table:EvolGame_parameter} and averaged over 100 independent simulation runs.} 
	\label{fig:primary} 
\end{figure*}

For the fixed population case, we solve Eq.~(\ref{eq:Nc_dot_detail}) numerically to estimate the stationary state values of $N_{C}$ and then computed the cooperation level $\rho_{c, \textnormal{fixed}}$ in the $(T, S)$ space based on Eq.~(\ref{eq:fraction_cooperation}), see Figure~\ref{fig:primary}(a). The figure shows that, for the H and SD games (i.e., $S > 0$), $\rho_{c, \textnormal{fixed}}$ is 1 even for small $N_{C,Z}$ (there is variation around $S = 0$). For the SH and PD games (i.e., $S < 0$), $\rho_{c, \textnormal{fixed}}$ increases as $N_{C,Z}$ increases but never reaches a positive value.

According to the setup described in Section~\ref{section:setup}, we estimate the average stationary state value of cooperation level for the emergency evacuation case (i.e., $\rho_{c, \textnormal{evac}}$), see Figure~\ref{fig:primary}(b). Similar to the fixed population case, $\rho_{c, \textnormal{evac}}$ is nearly 1 for most of the $(T, S)$ space of the H and SD games (i.e., $S > 0$) regardless of $N_{C,Z}$. However, in the SH and PD games (i.e., $S < 0$), $\rho_{c, \textnormal{evac}}$ approaches 1 as $N_{C, Z}$ increases, demonstrating a noticeable effect of increasing $N_{C,Z}$. In \ref{section:sensitivity}, we present the influence of selection intensity $\beta$ and sensory range $l_s$ on $\rho_{c, \textnormal{evac}}$.

We then compare the cooperation level estimated in the case of fixed population size ($\rho_{c, \textnormal{fixed}}$) and the numerical simulation ($\rho_{c, \textnormal{evac}}$), i.e., 
\begin{equation}\label{eq:cooperation_level_diff}
	\Delta \rho_{c} = \rho_{c, \textnormal{evac}}-\rho_{c, \textnormal{fixed}}.
\end{equation}
\noindent
Figure~\ref{fig:primary}(c) shows the cooperation level difference for various numbers of committed volunteers $N_{C,Z}$. When comparing the fixed population case to the emergency evacuation case, the cooperation level increases by up to 0.9 when $N_{C,Z}$ is large, while it reduces by up to 1.2 for small $N_{C,Z}$. When $N_{C,Z}$ is small, a negative difference is observed in the regions of H and SD games near the line $S = 0$, indicating that the evacuation process has a negative impact on the cooperation level. In contrast, if $N_{C,Z}$ is large, there is a positive difference through the SH and PD regions, implying that the cooperation level is further increased due to the evacuation process. 

\begin{figure*}[!t]
	\centering
	\begin{tabular}{c}
		\includegraphics[width=13.0cm]{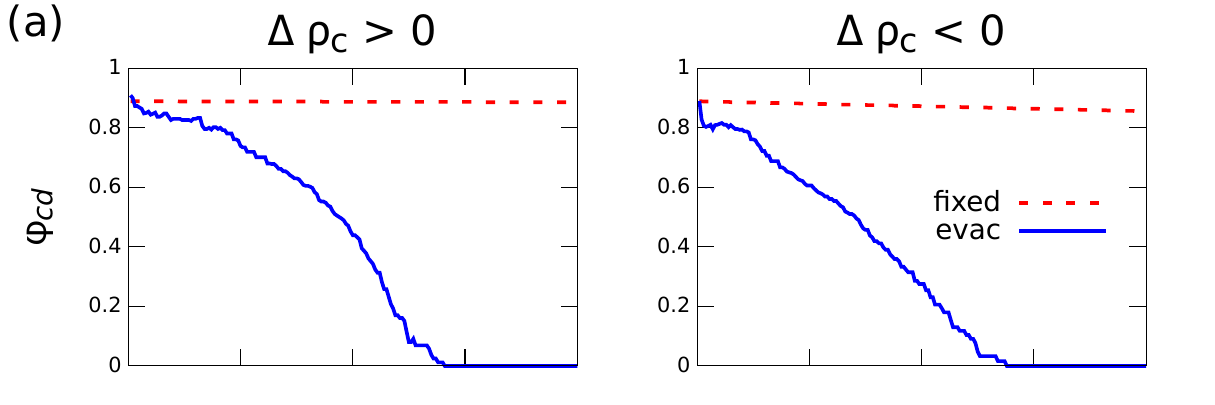}\vspace{-0.5cm}\\
		\includegraphics[width=13.0cm]{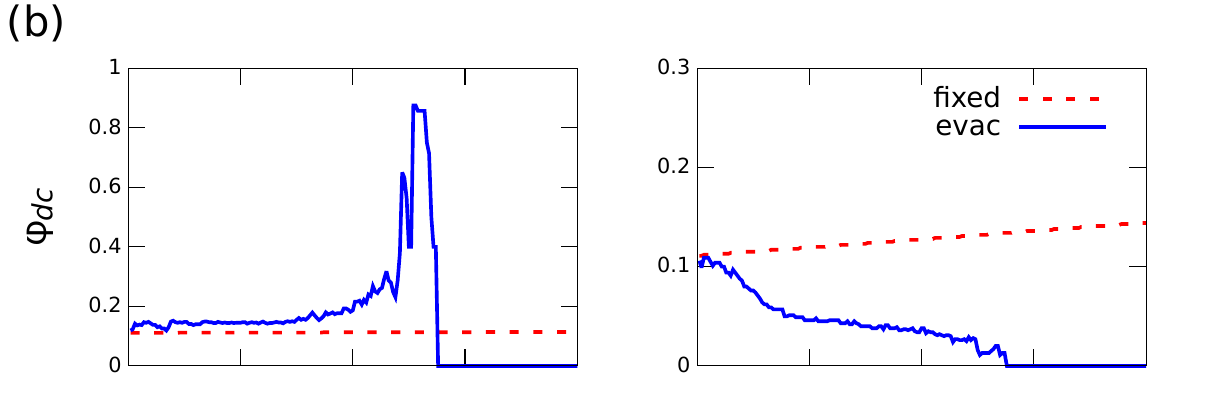}\vspace{-0.5cm}\\
		\includegraphics[width=13.0cm]{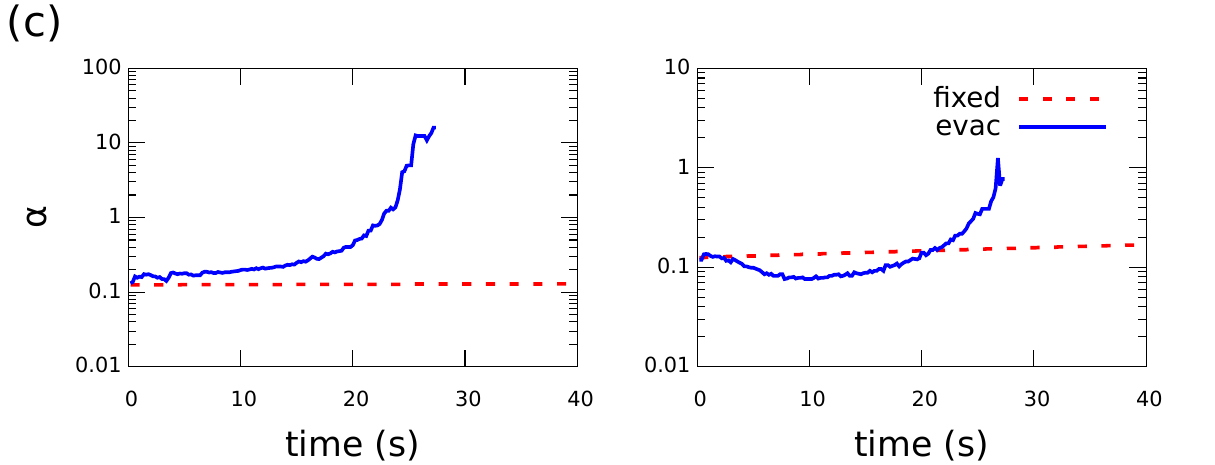}\vspace{-0.3cm}\\	
	\end{tabular}	
	\caption{Representative time courses of (a) the average fraction of defecting neighbors within the sensory range of a cooperator, i.e., $\phi_{cd}$, (b) the average fraction of cooperating neighbors within the sensory range of a defector, i.e., $\phi_{dc}$, and (c) the ratio of those fractions, i.e., $\alpha = \phi_{dc}/\phi_{cd}$. Results from the fixed population case are indicated by red dashed lines (i.e., $\alpha_{\textnormal{fixed}}$) and blue solid lines for the emergency evacuation case (i.e., $\alpha_{\textnormal{evac}}$). Shown in left and right columns are examples of $\Delta \rho_{c} > 0$ and $\Delta \rho_{c} < 0$, respectively: the left column graphs are generated with $T = 1.5$, $S = -0.2$, $\beta = 10$, and $N_{C, Z} = 10$, and the right column graphs are produced with  $T = 1.6$, $S = 0.2$, $\beta = 10$, and $N_{C, Z} = 2$. In the left column of (c), $\alpha_{\textnormal{evac}} > \alpha_{\textnormal{fixed}}$, indicating that more defectors tend to switch to cooperators in the emergency evacuation simulations than predicted by the case of fixed population. In the right column of (c), on the other hand, $\alpha_{\textnormal{evac}} < \alpha_{\textnormal{fixed}}$ for time $t < 22$~s, indicating that the evacuation process diminishes cooperation.}
	\label{fig:interaction_rate} 
\end{figure*}

To elucidate how the evacuation process impacts the cooperation level (see  Figure~\ref{fig:primary}(c)), we examine the fraction of neighbors susceptible to strategy adaptation per cooperator and defector. Analogous to connectedness as defined in the evolutionary dynamics of complex networks~\cite{Poncela_NJP2007, GomezGardenes_PRL2007}, we measure the average number of cooperating neighbors $M_{cc}(t)$ and defecting neighbors $M_{cd}(t)$ within the sensory range for each cooperator at time $t$. Based on $M_{cc}(t)$ and $M_{cd}(t)$, we define the fraction of neighbors susceptible to strategy adaptation from cooperation to defection at time $t$ as
\begin{equation}
	\label{eq:fraction_CD} 
	\phi_{cd}(t) = \frac{M_{cd}(t)}{M_{cd}(t)+M_{cc}(t)}.
\end{equation}
\noindent
Likewise, we also count the average number of cooperating neighbors $M_{dc}(t)$ and defecting neighbors $M_{dd}(t)$ within the sensory range for each defector, and measure the fraction of neighbors susceptible to strategy adaptation from defection to cooperation as
\begin{equation}
	\label{eq:fraction_DC} 
	\phi_{dc}(t) = \frac{M_{dc}(t)}{M_{dc}(t)+M_{dd}(t)}.
\end{equation}
\noindent
We then evaluate the ratio of $\phi_{dc}(t)$ and $\phi_{cd}(t)$, i.e.,
\begin{equation}\label{eq:ratio_fraction}
	\alpha = \frac{\phi_{dc}}{\phi_{cd}}.
\end{equation}
\noindent
When the value of $\alpha$ estimated in the emergency evacuation case ($\alpha_{\textnormal{evac}}$) is larger than that estimated in the fixed population case ($\alpha_{\textnormal{fixed}}$), i.e., $\alpha_{\textnormal{evac}} > \alpha_{\textnormal{fixed}}$, the emergency evacuation simulation yields higher cooperation level $\rho_{c}$ and consequently $\Delta \rho_{c} > 0$. Figure~\ref{fig:interaction_rate} illustrates representative time courses of $\phi_{dc}$, $\phi_{cd}$, and $\alpha$ evaluated for the fixed population and the emergency evacuation cases. 

For the emergency evacuation case, the number of defectors decreases over time due to the evacuation process. As indicated in the left column of Figure~\ref{fig:interaction_rate}(a), the fraction of defecting neighbors per cooperator $\phi_{cd}(t)$ decreases, leading to a decrease of neighbors who are susceptible to strategy adaptation from cooperation to defection. In the region of SH and PD games with large $N_{C,Z}$, the influence of negative $S$ on the cooperator's payoff (i.e., $u_C(t)~\sim~R\left\langle M_{cc} \right\rangle(t)+S\left\langle M_{cd} \right\rangle(t)$) becomes less severe, possibly increasing the number of cooperators. In addition, larger $N_{C,Z}$ tends to reduce $\phi_{cd}(t)$ further, so the cooperator's payoff may increase further. At the same time, the fraction of cooperating neighbors per defector $\phi_{dc}(t)$ grows, seemingly due to the increased number of cooperators. However, $\phi_{dc}(t)$ in the fixed population case is smaller than that in the emergency evacuation case [see the left column of Figure~\ref{fig:interaction_rate}(b)]. From the left column of Figure~\ref{fig:interaction_rate}(c), one can find that $\alpha_{\textnormal{evac}}$ (blue solid line) is larger than $\alpha_{\textnormal{fixed}}$ (red dashed line), indicating that more defectors tend to switch to cooperators in the emergency evacuation simulations than predicted by the case of fixed population. This results in further enhancement of cooperation as shown in the panel of $N_{C,Z} = 10$ in Figure~\ref{fig:primary}(c), yielding $\Delta \rho_{c} > 0$. In the region of H and SD games with small $N_{C,Z}$, the value of $S$ is nearly 0 or positive, thus a decrease in $\phi_{cd}(t)$ potentially reduces the cooperator's payoff and the number of cooperators. At the same time, the fraction of cooperating neighbors per defector $\phi_{dc}(t)$ decreases and becomes smaller than that in the fixed population case, see the right column of Figure~\ref{fig:interaction_rate}(b). In the right column of Figure~\ref{fig:interaction_rate}(c), the value of $\alpha_{\textnormal{evac}}$ (blue solid line) is smaller than $\alpha_{\textnormal{fixed}}$ (red dashed line) for time $t < 22$~s, indicating that the evacuation process diminishes cooperation (i.e., $\Delta \rho_{c} < 0$).

\subsection{Phase diagrams}
\label{section:Results_PhaseDiagrams}

\begin{figure*}[!t]
	\centering
	\begin{tabular}{c}
		\includegraphics[width=15.5cm]{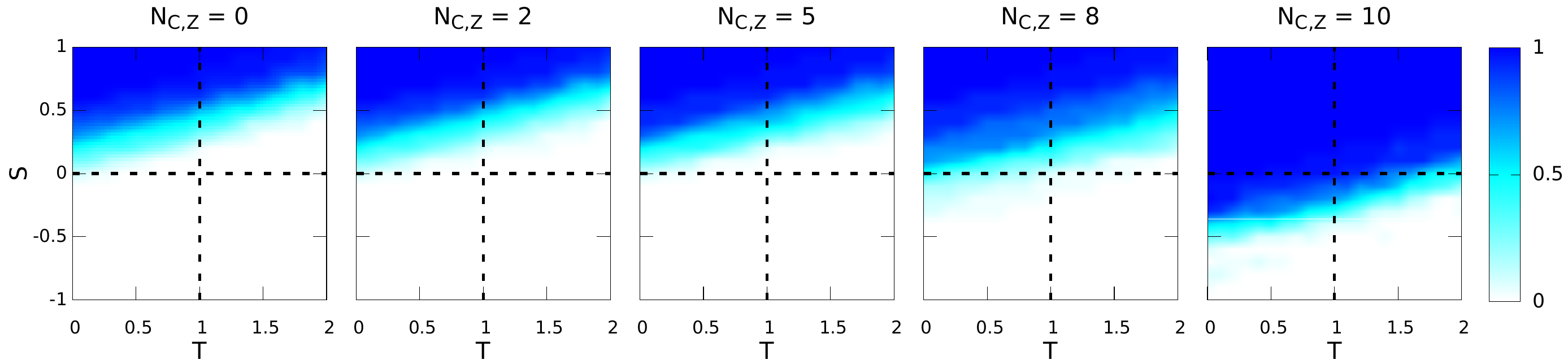}\vspace{-0.3cm}\\
	\end{tabular}	
	\caption{The complete rescue probability $P_r$ of the emergency evacuation case with different numbers of committed volunteers $N_{C, Z}$. Blue area in each panel expands as $N_{C, Z} > 5$ increases, implying that $P_r$ tends to grow. Results are generated with parameter values listed in Table~\ref{table:EvolGame_parameter} and averaged over 100 independent simulation runs.} 
	\label{fig:rescue_prob} 
\end{figure*}

In the context of emergency evacuations, it is interesting to examine whether all the injured persons are rescued. We measure the complete rescue probability $P_r$ by counting the occurrence of complete rescue over 100 independent simulation runs for each parameter combination $(T, S)$ along with a given value of $N_{C,Z}$. Figure~\ref{fig:rescue_prob} presents the general tendency of the complete rescue probability $P_r$. For a given value of $(T, S)$, $P_r$ tends to grow as $N_{C,Z}$ increases.

\begin{figure*}[!t]
	\centering
	\begin{tabular}{c}
		\includegraphics[width=14.0cm]{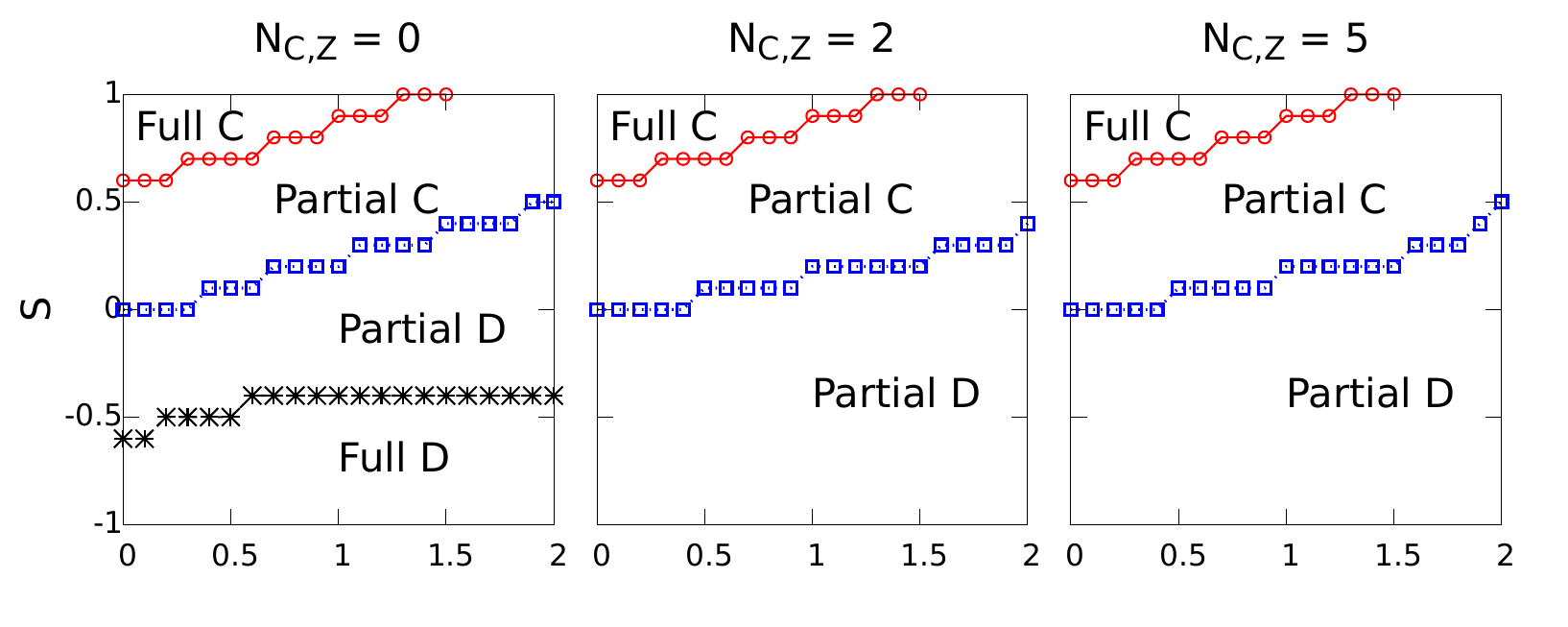}\vspace{-0.5cm}\\
		\includegraphics[width=14.0cm]{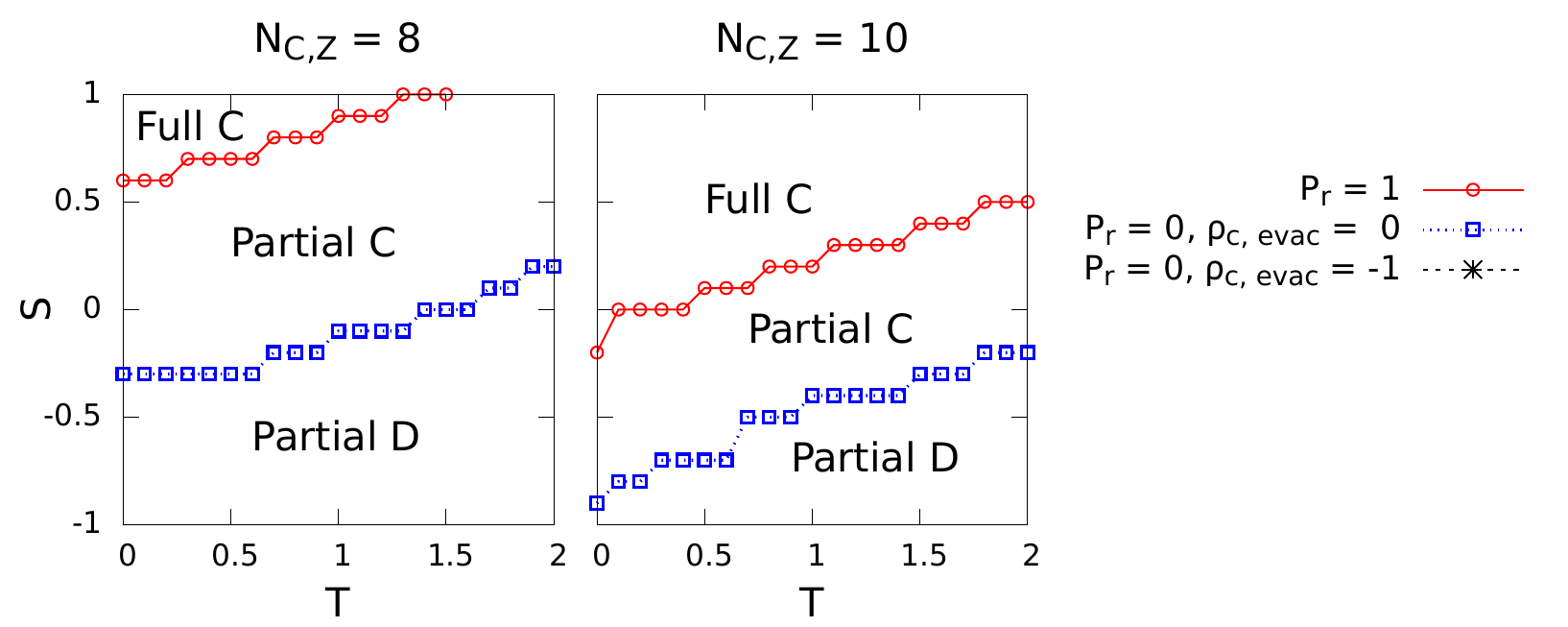}\vspace{-0.5cm}\\
	\end{tabular}	
	\caption{Phase diagrams summarizing the numerical results. The $(T, S)$ space is divided into different phases by means of the complete rescue probability $P_r$ and the cooperation level $\rho_{c, \textnormal{evac}}$: full cooperation ($P_r = 1$), partial cooperation ($1 > P_r > 0$ and $1 > \rho_{c, \textnormal{evac}} > 0$), partial defection ($P_r = 0$ and $0 > \rho_{c, \textnormal{evac}} > -1$), and full defection ($P_r = 0$ and $\rho_{c, \textnormal{evac}} = -1$) phases. Different symbols represent the boundaries between different phases: red circles ($\bigcirc$) for the boundary between the full cooperation and the partial cooperation phases ($P_r = 1$), blue rectangles ($\square$) between the partial cooperation and the partial defection phases ($P_r = 0$ and $\rho_{c, \textnormal{evac}} = 0$), and black asterisks ($\ast$) between the partial defection and the full defection phases ($P_r = 0$ and $\rho_{c, \textnormal{evac}} = -1$). Here, Full C, Partial C, Partial D, and Full D indicate the full cooperation phase, partial cooperation phase, partial defection phase, and full defection phase, respectively. One can notice that, for large $N_{C,Z}$, the boundary of the full and partial cooperation phases expand downward in the $(T, S)$ space as $N_{C,Z}$ increases. Results are generated with parameter values listed in Table~\ref{table:EvolGame_parameter} and averaged over 100 independent simulation runs.}
	\label{fig:phases} 
\end{figure*}

Based on the cooperation level $\rho_{c, \textnormal{evac}}$ and the complete rescue probability $P_r$, we characterize different phases of collective helping behavior in the $(T, S)$ space: the {\em full cooperation}, {\em partial cooperation}, {\em partial defection}, and {\em full defection phases}. We define the full cooperation phase by $P_r = 1$. In this phase, all the lonely volunteers successfully find peer volunteers, thus all the injured are rescued. It is also noted that, according to our observations, $\rho_{c, \textnormal{evac}}$ is always 1 for $P_r = 1$. The partial cooperation phase is characterized by $1 > P_r > 0$ and $1 > \rho_{c, \textnormal{evac}} > 0$, in which the complete rescue is not always possible seemingly because the number of volunteers in the stationary state is changing depending on the randomness. The partial defection phase is defined by $P_r = 0$ and $0 > \rho_{c, \textnormal{evac}} > -1$, implying that some lonely volunteers turn into bystanders thus some injured persons are not rescued. We characterize the full defection phase by $P_r = 0$ and $\rho_{c, \textnormal{evac}} = -1$, suggesting that none of the injured are rescued because all the lonely volunteers turn into bystanders. Figure~\ref{fig:phases} summarizes the phase characterization of our emergency evacuation simulation results. 

\begin{figure*}[!t]
	\centering
	\begin{tabular}{c}
		\includegraphics[width=9.0cm]{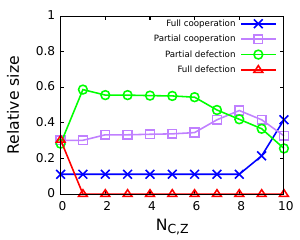}\vspace{-0.3cm}\\
	\end{tabular}	
	\caption{Relative size of different phases in the $(T, S)$ space. At $N_{C,Z} = 6$, the relative size of partial defection phase begins to decrease while that of partial cooperation increases, suggesting that adding additional committed volunteers is effective in promoting cooperation only for $N_{C,Z} \ge 6$. Different symbols represent the results of different phases: blue cross ($\times$) for the full cooperation phase, purple rectangle ($\square$) for the partial cooperation phase, green circle ($\bigcirc$) for the partial defection, and red triangle ($\bigtriangleup$) for the full defection phase. Results are generated with parameter values listed in Table~\ref{table:EvolGame_parameter}.}
	\label{fig:RelativePhaseSize} 
\end{figure*}

In addition to the characterization of different collective helping behavior in phase diagrams, we evaluate the relative size different phases in the $(T, S)$ space by counting the number of $(T, S)$ parameter combinations for each phase. As can be seen from Figure~\ref{fig:RelativePhaseSize}, the full defection phase disappears when the number of committed volunteers $N_{C,Z}$ is larger than 0, indicating that the existence of committed volunteers prevents the situation in which none of the injured are rescued. For $1 < N_{C,Z} < 6$, the relative size of each phase is virtually constant, implying that increasing the number of committed volunteers is not beneficial to the spread of cooperation. For $N_{C,Z} > 6$, a larger number of committed volunteers contributes to the spread of cooperation. In particular, when $N_{C,Z} > 8$, the relative size of the full cooperation phase is significantly increased, suggesting that a complete rescue is more frequently observed. Note that the behavior of the relative phase size curve is different depending on the values of selection intensity $\beta$ and sensory range $l_s$, refer to Figure~\ref{fig:RelativePhaseSize_beta} for the influence of $\beta$ and Figure~\ref{fig:RelativePhaseSize_ls} for the influence of $l_s$.

\section{Conclusion}
\label{section:conclusion}

In this study, we have developed an evolutionary game theoretic model to study the helping behavior in emergency evacuations with the existence of committed volunteers. By systematically controlling the number of committed volunteers and payoff parameters, we have shown how the presence of committed volunteers influence the collective helping behavior based on the cooperation level $\rho_{c}$ and complete rescue probability $P_{r}$. We have observed that committed volunteers can facilitate the spread of cooperation in the population. However, adding additional committed volunteers is effective only above a certain number of committed volunteers, analogous to the threshold effect reported in literature~\cite{Cardillo_PRR2020}. Interestingly, unlike in well-mixed populations with a fixed number of individuals, the evacuation process might further enhance the cooperation level for the case of small suckers' payoff and larger numbers of committed volunteers. This can be attributed to the change in fractions of neighbors who are susceptible to strategy adaptation. 

There are a number of possible directions for future work addressing the limitations of this study. To focus on the essential features of the strategic interactions in helping behavior, this study has considered a simple room evacuation scenario. The presented model needs to be tested with more complex geometry conditions to further examine the complex nature of strategic interactions among individuals. In this study, we also assumed that the control parameter values are the same for all individuals, but investigating the impact of heterogeneity in individuals' characteristics would be an interesting avenue to explore. Examples include mutations~\cite{Amaral_PRE2018}, different interaction rules among individuals~\cite{Mobilia_PRE2012, Woelfing_JTB2009,  Shigaki_PRE2012, Su_NJP2016}, and various payoff parameter setups for different individuals~\cite{Hashimoto_JTB2014, Amaral_PRE2016a}. 

In future studies, the presented model can be verified against computer-based experiments in virtual environments~\cite{Bode_RSOS2015, Kinateder_AppErgonomics2014, Lovreglio_SafetySci2020}. Such computer-based experiments have applied to study human behavior in emergency evacuations like human route choice behavior~\cite{Bode_RSOS2015} and the influence of conflicting information~\cite{Kinateder_AppErgonomics2014}. By means of computer-based experiments, real pedestrian behaviors can be observed and tested for various emergency situations that can be difficult or even risky to perform direct test of real pedestrians. 

Another possible direction of future work could be to examine the effects of network structures~\cite{Cardillo_PRR2020, Shigaki_PRE2012, Matsuzawa_PRE2016, Li_NatureComm2020} on the evolutionary game dynamics. The game theoretic model presented in this paper can be extended to predict the evolution of cooperation in pedestrian movement (for instance, exit selection in evacuation~\cite{Heliovaara_SafetySci2012}) by analyzing the underlying interaction network structures characterized from pedestrian trajectories.

\section*{Acknowledgements}
This research is supported by National Research Foundation (NRF) Singapore, GOVTECH under its Virtual Singapore Program Grant No. NRF2017VSG-AT3DCM001-031. We thank Mr. Vinayak Teoh Kannappan for his help in implementing the simulation model presented in Section~\ref{section:setup} with C++.

\section*{Data availability statement}
Data sharing is not applicable to this article as no new data were created or analyzed in this study. The source code for the helping behavior model can be found at the Github page: \url{https://github.com/jaeyoung82/PedEvacHelpingBehavior}. 

\appendix
\section{Details of the social force model}
\label{section:SFM_details}

In Section~\ref{section:SFM}, we presented a general form of the social force models~\cite{Helbing_PRE1995,Helbing_RMP2001,Johansson_ACS2007}. This appendix provides further details of the presented social force model.

The social force models describe the acceleration of pedestrian $i$ as a superposition of driving and repulsive force terms according to the following equation of motion:
\begin{equation}\label{eq:EoM}
	\frac{d\vec{v}_i(t)}{dt} = \vec{f}_{i,d}+\sum_{j\neq i}^{ }{\vec{f}_{ij}}+\sum_{B}^{ }{\vec{f}_{iB}}.
\end{equation}

The driving force $\vec{f}_{i,d}$ is given as
\begin{equation}\label{eq:driving}
	\vec{f}_{i,d} = \frac{v_d\vec{e}_{i}-\vec{v}_{i}(t)}{\tau},
\end{equation}
\noindent
where ${v}_{d}$ is the desired speed and $\vec{e}_{i}$ is a unit vector indicating the desired walking direction of pedestrian $i$. The relaxation time $\tau$ controls how fast the pedestrian $i$ adapts its velocity to the desired velocity. 

The interpersonal repulsive force term $\vec{f}_{ij}$ is specified according to the circular specification~\cite{Helbing_RMP2001} which is the most simplistic form of the interpersonal repulsive interaction. The explicit form of $\vec{f}_{ij}$ can be written as 
\begin{equation}\label{eq:CS}
	\vec{f}_{ij} = C_p exp \left(\frac{r_i+r_j-d_{ij}}{l_p}\right) \vec{e}_{ij},
\end{equation}
\noindent
where $\vec{e}_{ij} = \vec d_{ij}/d_{ij}$ is a unit vector pointing from pedestrian $j$ to pedestrian $i$, and $\vec d_{ij} = \vec x_i-\vec x_j$ is the distance vector pointing from pedestrian~$j$ to pedestrian~$i$. The strength and the range of repulsive interaction between pedestrians are denoted by $C_p$ and $l_p$, respectively. Instead of the circular specification, one might select a different form of specifications such as elliptical specification I (ES-1)~\cite{Helbing_PRE1995} and elliptical specification II (ES-2)~\cite{Johansson_ACS2007}. 

In addition to the interpersonal repulsive force term $\vec{f}_{ij}$ presented in Eq.~(\ref{eq:CS}), the interpersonal elastic force term $\vec g_{ij}$ is added when the distance $d_{ij}$ is smaller than the sum $r_{ij} = r_{i}+r_{j}$ of their radii $r_{i}$ and $r_{j}$. We describe $\vec g_{ij}$ as Helbing~\cite{Helbing_RMP2001} suggested, 
\begin{equation}\label{eq:friction_full}
	\vec{g}_{ij} = h(r_{ij}-d_{ij}) \left\{k_n \vec{e}_{ij}+k_t [(\vec{v}_j-\vec v_i)\cdot \vec{t}_{ij}]\vec{t}_{ij}\right\},
\end{equation}
\noindent
where $k_n$ and $k_t$ are the normal and tangential elastic constants, respectively. A unit vector $\vec{e}_{ij}$ is pointing from pedestrian $j$ to pedestrian $i$, and $\vec{t}_{ij}$ is a unit vector perpendicular to $\vec{e}_{ij}$. The function $h(x)$ yields $x$ if $x > 0$, while it gives $0$ if $x \leq 0$. 

The repulsive force from boundaries $\vec{f}_{iB}$ is given as 
\begin{equation}\label{eq:boundary}
	\vec{f}_{iB} = C_b\exp\left(-\frac{d_{iB}}{l_{b}}\right)\vec{e}_{iB},
\end{equation}
\noindent
where $d_{iB}$ is the perpendicular distance between pedestrian $i$ and wall, and $\vec{e}_{iB}$ is the unit vector pointing from the wall $B$ to the pedestrian $i$. The strength and range of repulsive interaction from boundaries are denoted by $C_b$ and $l_b$. 

\begin{table}
	\centering
	\caption{\label{table:SFM_parameter}Social force model parameters}
	\begin{indented}
		\lineup
		\item[]
		\begin{tabular}{@{}lll}
			\br
			Model parameter & symbol & value \\
			\mr
			interpersonal repulsion strength	& $C_p$ & 3.0 \\
			interpersonal repulsion range 		& $l_p$ & 0.2 \\
			normal elastic constants 			& $k_n$ & 50.0 \\
			tangential elastic constants 		& $k_t$ & 25.0 \\
			boundary repulsion strength 		& $C_b$ & 10.0 \\
			boundary repulsion range 			& $l_b$ & 0.1 \\
			\br
		\end{tabular}
	\end{indented}
\end{table}

To implement the presented social force model, the parameter values in Table~\ref{table:SFM_parameter} are selected based on the previous studies~\cite{Helbing_PRE1995, Helbing_RMP2001, Kwak_PRE2017, Kwak_PRE2013}.

The ambulant pedestrians move with the initial desired speed $v_d = v_{d, 0}=1.2$~m/s and with relaxation time $\tau = 0.5$~s, and their speed cannot exceed $v_{\rm max} = 2.0$~m/s. Until now, the speed of volunteers rescuing the injured persons is often assumed by the modelers, like the work of von~Sivers~\textit{et al.}~\cite{vonSivers_PED2014,vonSivers_SafetySci2016}. We applied speed reduction factor $\alpha = 0.5$ to the volunteers rescuing the injured persons, so they move with a reduced desired speed $v_{d} = \alpha v_{d, 0} = 0.6$~m/s. Following previous studies~\cite{Kwak_PRE2017, Zanlungo_EPL2011, Zanlungo_PRE2014}, we discretized the numerical integration of Eq.~(\ref{eq:EoM}) using the first-order Euler method:
\begin{eqnarray}\label{eq:Euler_method}
	\vec{v}_i(t + \Delta t) &=& \vec{v}_i(t) + \vec{a}_i(t)\Delta t,\\
	\vec{x}_i(t + \Delta t) &=& \vec{x}_i(t) + \vec{v}_i(t + \Delta t)\Delta t.
\end{eqnarray}
\noindent
Here, $\vec{a}_i(t)$ is the acceleration of pedestrian $i$ at time $t$ which can be obtained from Eq.~(\ref{eq:EoM}). The velocity and position of pedestrian $i$ is denoted by $\vec{v}_i(t)$ and $\vec{x}_i(t)$, respectively. The time step $\Delta t$ is set as 0.05~s. 

\section{Sensitivity analysis}
\label{section:sensitivity}
We conducted sensitivity analysis since the general tendency of collective helping behavior presented in Section~\ref{section:Results_General} is sensitive to the parameter values listed in Table~\ref{table:EvolGame_parameter}, particularly the selection intensity $\beta$ and the sensory range $l_s$. 

\begin{figure*}[!t]
	\centering
	\begin{tabular}{c}
		\includegraphics[width=12.0cm]{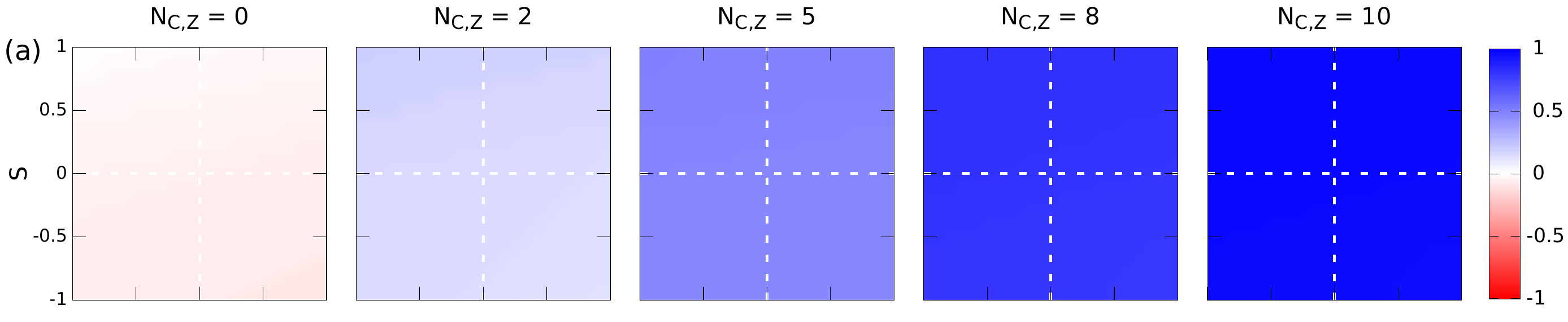}\vspace{-0.3cm}\\
		\includegraphics[width=12.0cm]{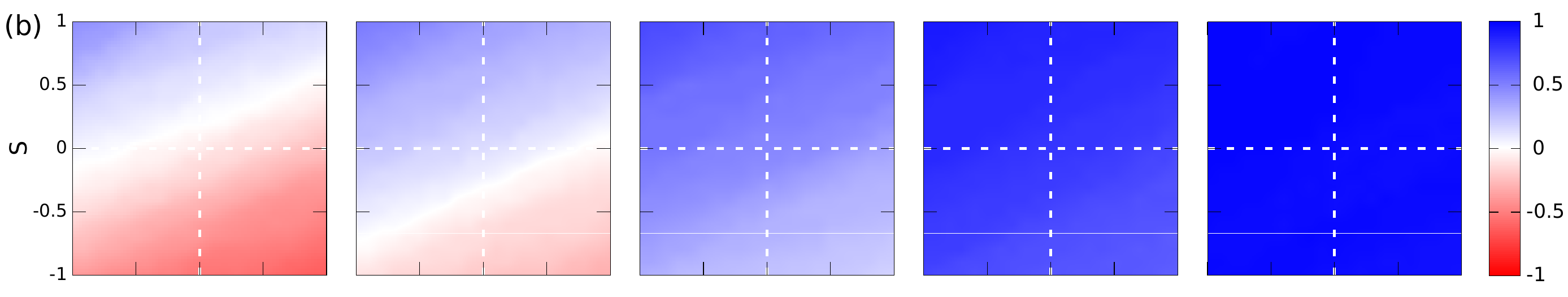}\vspace{-0.3cm}\\
		\includegraphics[width=12.0cm]{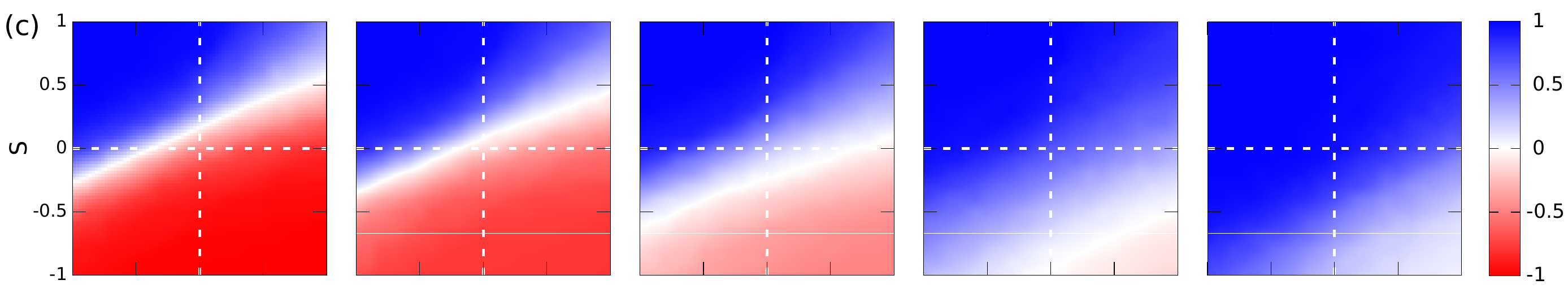}\vspace{-0.3cm}\\
		\includegraphics[width=12.0cm]{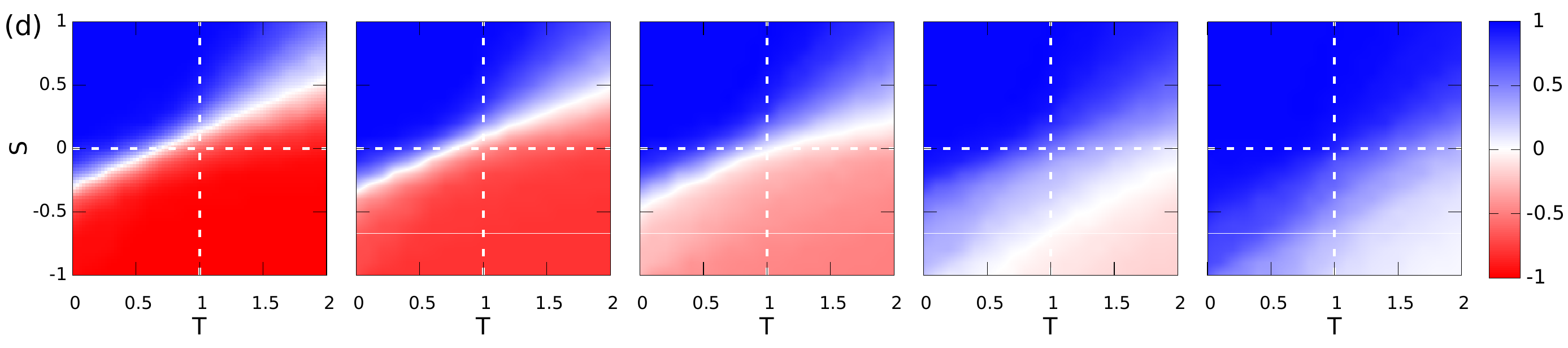}\vspace{-0.3cm}\\
	\end{tabular}	
	\caption{The average stationary state value of cooperation level $\rho_{c, \textnormal{evac}}$ of the emergency evacuation case with different values of selection intensity $\beta$: (a) $\beta = 0.1$, (b) $\beta = 1$, (c) $\beta = 10$, and (d) $\beta = 100$. Each panel corresponds to different numbers of committed volunteers $N_{C, Z}$. Changing values of $T$ and $S$ can make difference in $\rho_{c, \textnormal{evac}}$ as $\beta$ increases, but it does not for small $\beta$. Results are averaged over 100 independent simulation runs.} 
	\label{fig:T-S_beta} 
\end{figure*}

\begin{figure*}[!t]
	\centering
	\begin{tabular}{c}
		\includegraphics[width=12.0cm]{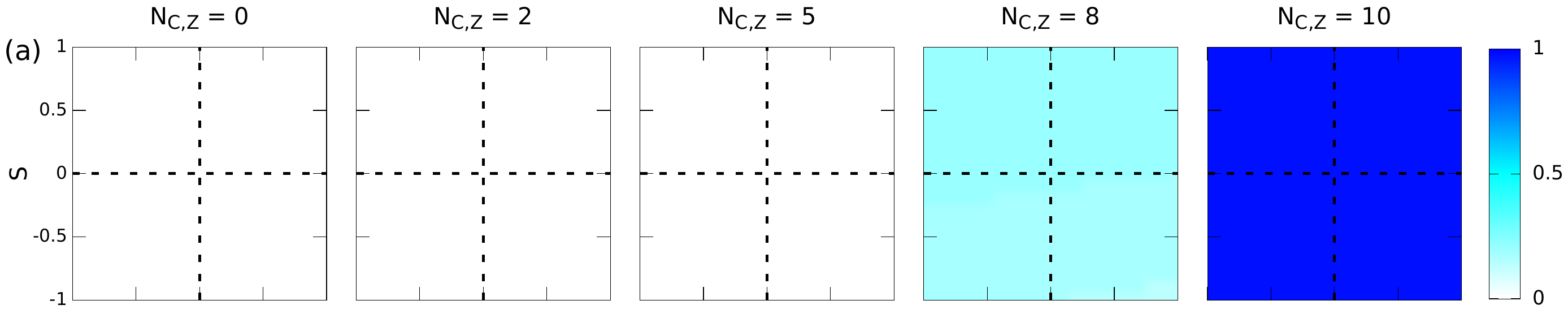}\vspace{-0.3cm}\\
		\includegraphics[width=12.0cm]{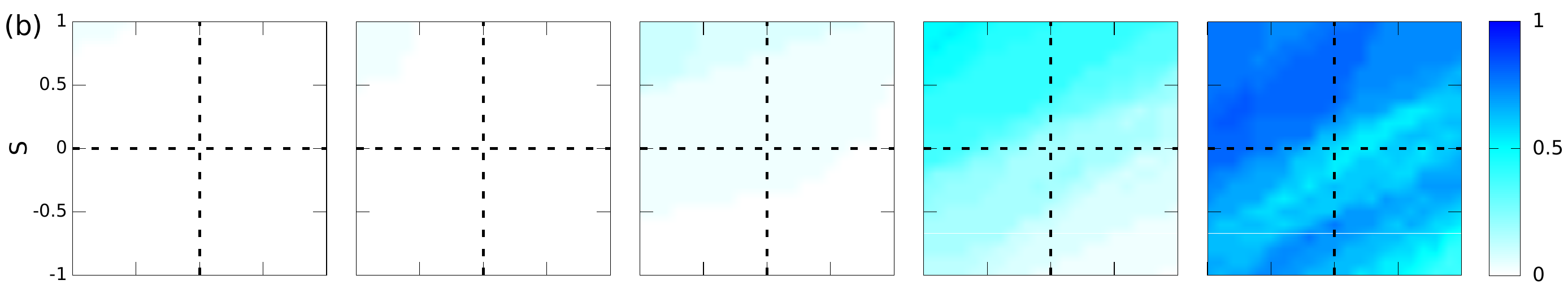}\vspace{-0.3cm}\\
		\includegraphics[width=12.0cm]{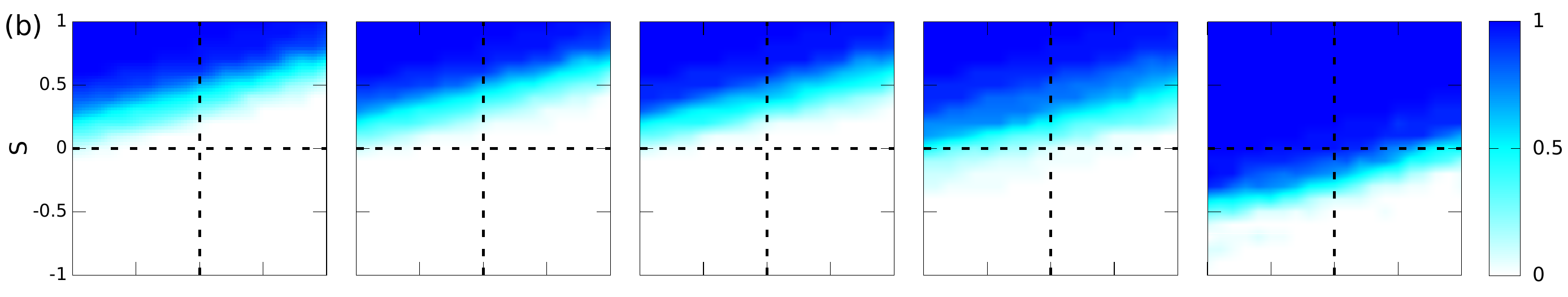}\vspace{-0.3cm}\\
		\includegraphics[width=12.0cm]{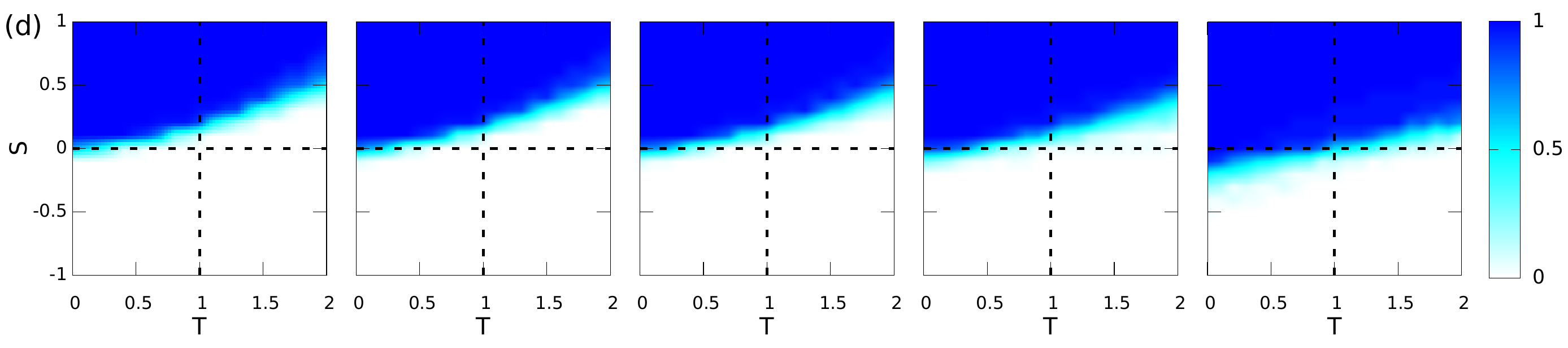}\vspace{-0.3cm}\\
	\end{tabular}	
	\caption{The complete rescue probability $P_r$ of the emergency evacuation case with different values of selection intensity $\beta$: (a) $\beta = 0.1$, (b) $\beta = 1$, (c) $\beta = 10$, and (d) $\beta = 100$. Each panel corresponds to different numbers of committed volunteers $N_{C, Z}$. For small $\beta$, the region of $P_r \approx 1$ is growing as $N_{C,Z}$ increases. However, the effect of $N_{C,Z}$ on the growth of $P_r \approx 1$ region is little for large $\beta$. Results are averaged over 100 independent simulation runs.} 
	\label{fig:Pr_beta} 
\end{figure*}

\begin{figure*}[!t]
	\centering
	\begin{tabular}{c}
		\includegraphics[width=15.5cm]{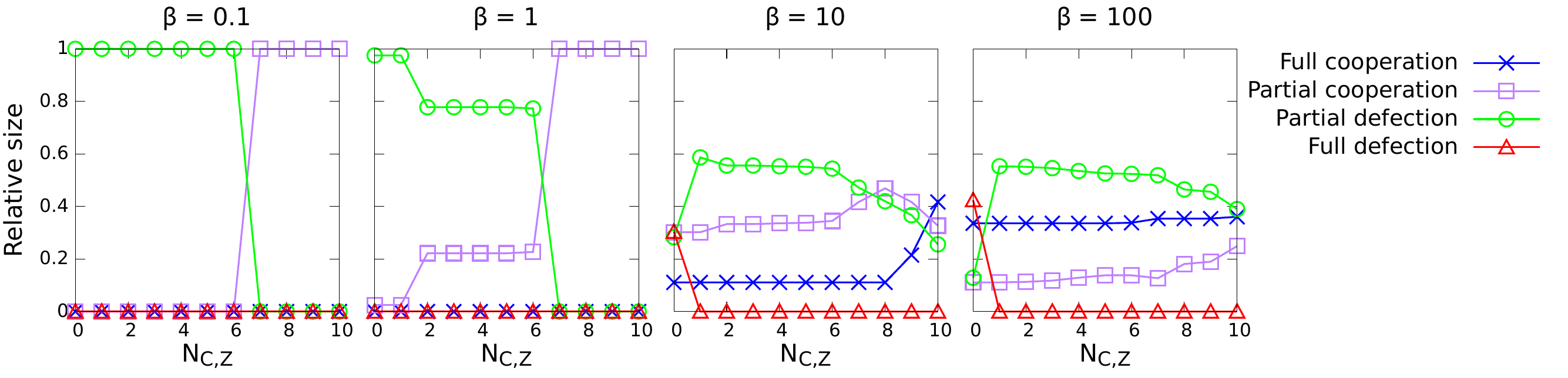}\vspace{-0.3cm}\\
	\end{tabular}	
	\caption{Relative size of different phases in the $(T, S)$ space for different values of selection intensity $\beta$: blue cross ($\times$) for the full cooperation phase, purple rectangle ($\square$) for the partial cooperation phase, green circle ($\bigcirc$) for the partial defection, and red triangle ($\bigtriangleup$) for the full defection phase. For $\beta \leq 1$, only the partial defection and partial cooperation phases exist. When $\beta > 1$, the full cooperation and full defection phases appear.}
	\label{fig:RelativePhaseSize_beta} 
\end{figure*}

Varying $\beta$ in our simulation yields substantial changes in collective helping behavior. As can be seen from Figure~\ref{fig:T-S_beta}, for a given value of the number of committed volunteers $N_{C,Z}$, increasing $\beta$ obliviously changes the behavior of $\rho_{c, \textnormal{evac}}$ in the $(T, S)$ space. For $\beta = 0.1$, $\rho_{c, \textnormal{evac}}$ is insensitive to the change of $T$ and $S$. As $\beta$ increases, different values of $T$ and $S$ make difference in $\rho_{c, \textnormal{evac}}$, showing good agreement with previous studies. Figure~\ref{fig:Pr_beta} presents the influence of $\beta$ on the complete rescue probability $P_r$. When $\beta$ is small, one can clearly see that the region of $P_r \approx 1$ is growing as $N_{C,Z}$ increases. However, when $\beta$ is large, the effect of $N_{C,Z}$ on the growth of $P_r \approx 1$ region is little. Figure~\ref{fig:RelativePhaseSize_beta} show how the collective helping behavior depend on $\beta$. If $\beta = 1$, one can observe either the partial defection or partial cooperation phases for a given value of $N_{C,Z}$. When $\beta \approx 1$, at least two different phases can be seen from the $(T, S)$ space for certain values of $N_{C,Z}$. For $\beta > 1$, the full cooperation and full defection phases appear. The effect of $N_{C,Z}$ is seemingly noticeable when $\beta$ is near 10, but the effect is getting less significant as we increase $\beta$ further.

\begin{figure*}[!t]
	\centering
	\begin{tabular}{c}
		\includegraphics[width=12.0cm]{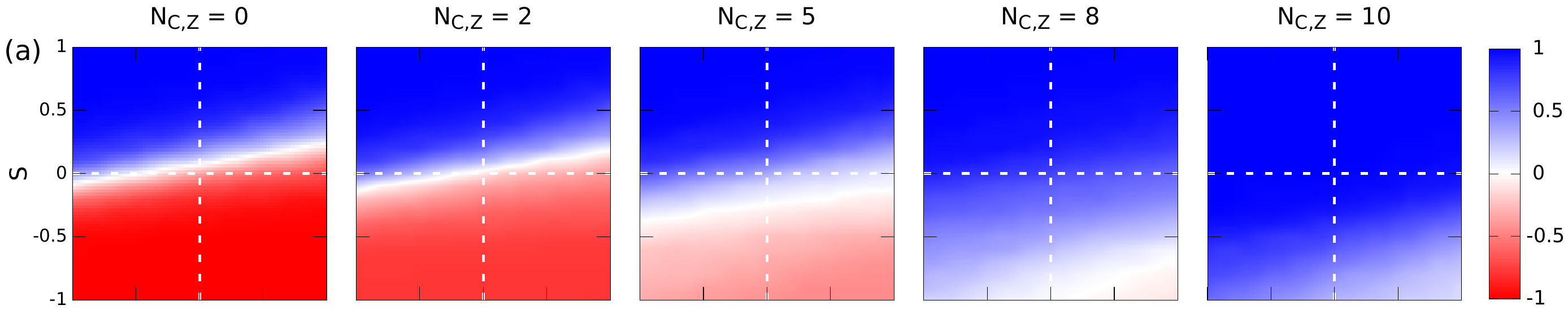}\vspace{-0.3cm}\\
		\includegraphics[width=12.0cm]{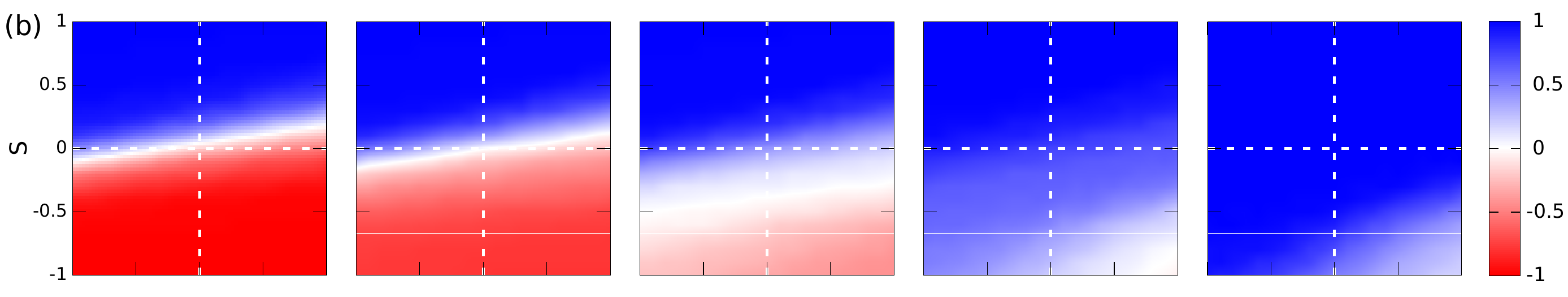}\vspace{-0.3cm}\\
		\includegraphics[width=12.0cm]{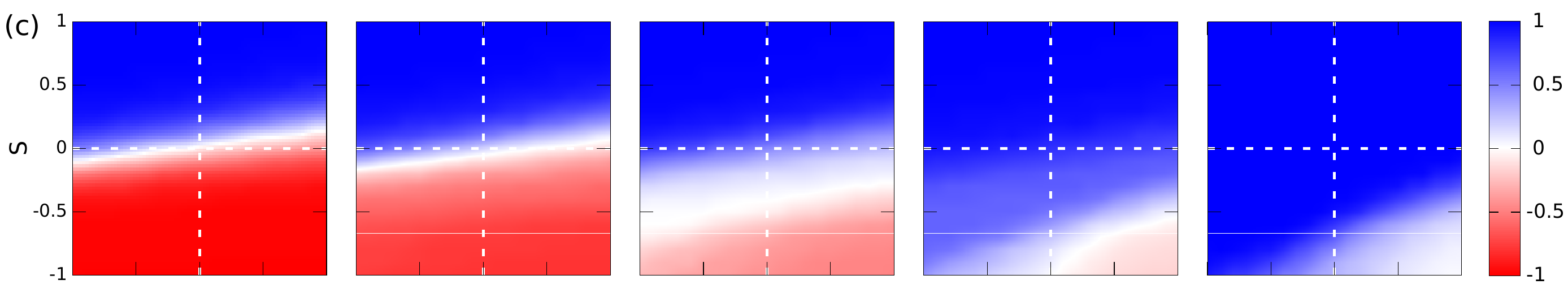}\vspace{-0.3cm}\\
		\includegraphics[width=12.0cm]{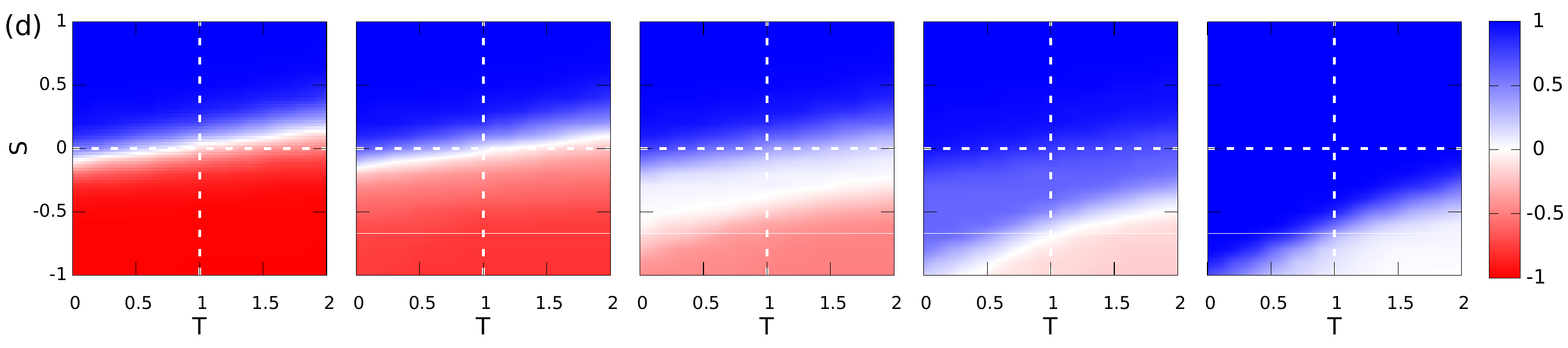}\vspace{-0.3cm}\\
	\end{tabular}	
	\caption{The average stationary state value of cooperation level $\rho_{c, \textnormal{evac}}$ of the emergency evacuation case with different values of sensory range $l_s$: (a) $l_s = 3$~m, (b) $l_s = 6$~m, (c) $l_s = 9$~m, and (d) $l_s = 15$~m. Each panel corresponds to different numbers of committed volunteers $N_{C, Z}$. Increasing $l_s$ from 3~m to 6~m shows notable increment in $\rho_{c, \textnormal{evac}}$ for high $N_{C, Z}$, but increasing $l_s$ further does not seem to make significant changes. Results are averaged over 100 independent simulation runs.} 
	\label{fig:T-S_ls} 
\end{figure*}

\begin{figure*}[!t]
	\centering
	\begin{tabular}{c}
		\includegraphics[width=12.0cm]{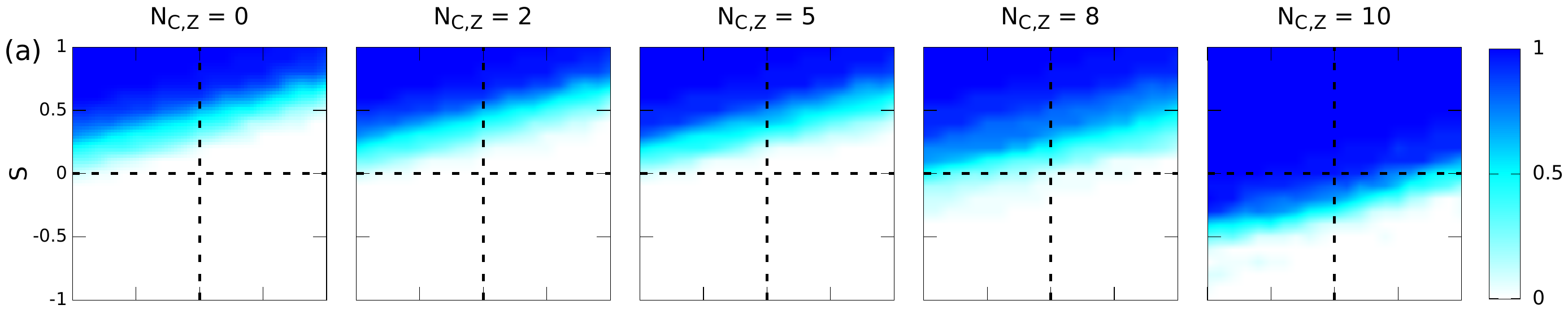}\vspace{-0.3cm}\\
		\includegraphics[width=12.0cm]{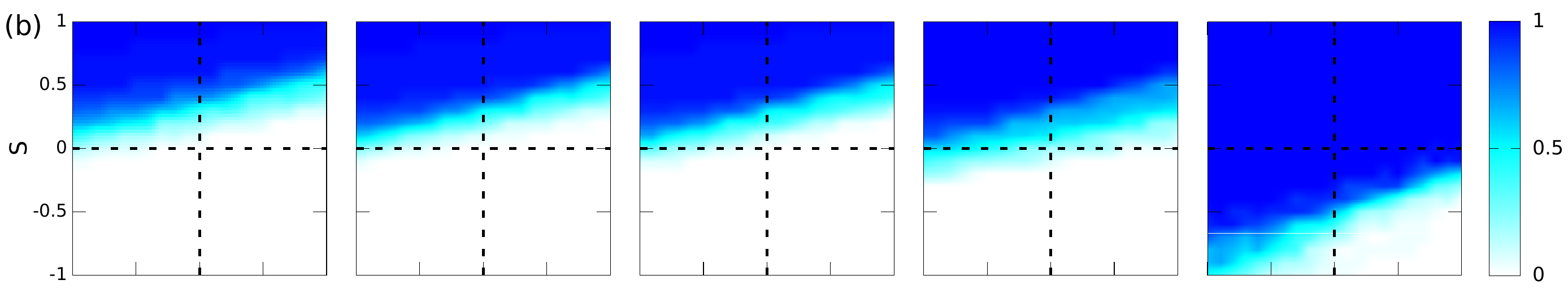}\vspace{-0.3cm}\\
		\includegraphics[width=12.0cm]{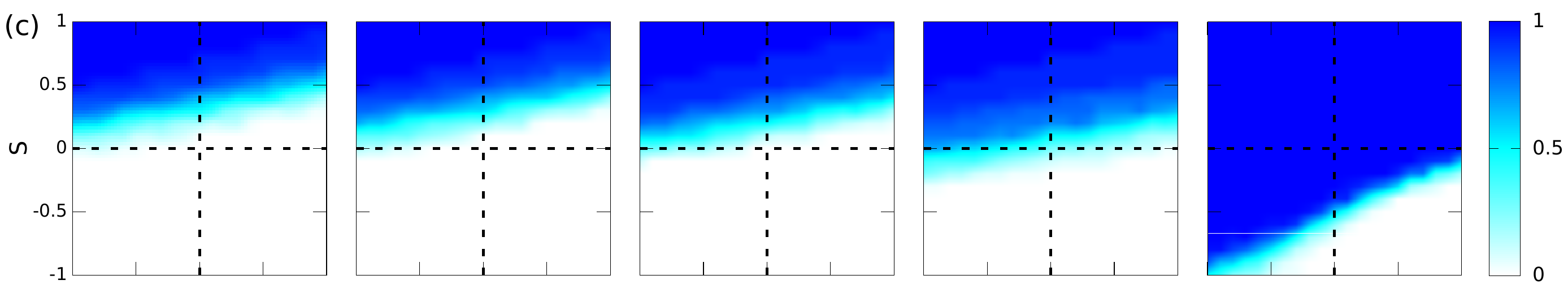}\vspace{-0.3cm}\\
		\includegraphics[width=12.0cm]{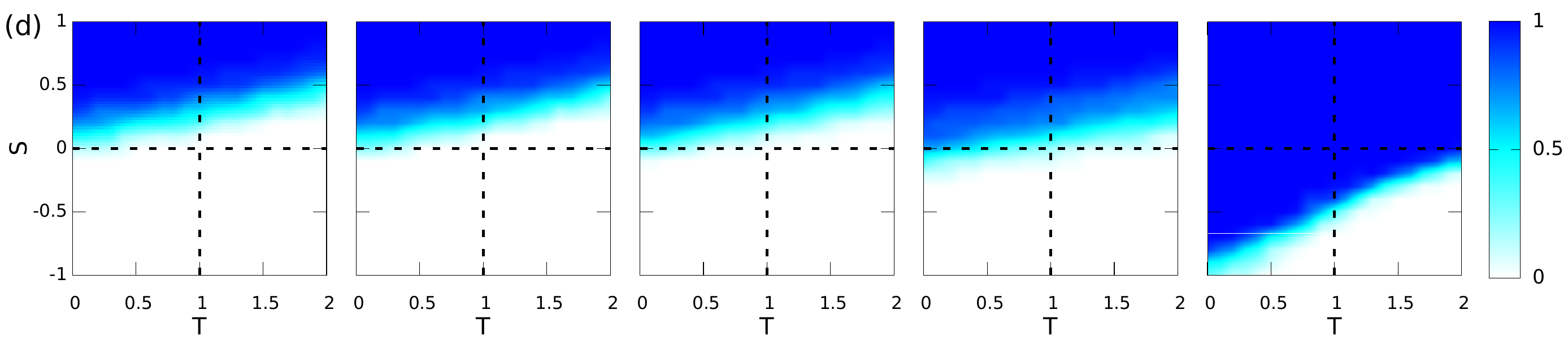}\vspace{-0.3cm}\\
	\end{tabular}	
	\caption{The complete rescue probability $P_r$ of the emergency evacuation case with different values of sensory range $l_s$: (a) $l_s = 3$~m, (b) $l_s = 6$~m, (c) $l_s = 9$~m, and (d) $l_s = 15$~m. Each panel corresponds to different numbers of committed volunteers $N_{C, Z}$. For high $N_{C, Z}$, increasing $l_s$ from 3~m to 6~m makes considerable changes in $P_r$, but it does not appear to be notable if $l_s$ is larger than 6~m. Results are averaged over 100 independent simulation runs.} 
	\label{fig:Pr_ls} 
\end{figure*}

\begin{figure*}[!t]
	\centering
	\begin{tabular}{c}
		\includegraphics[width=15.5cm]{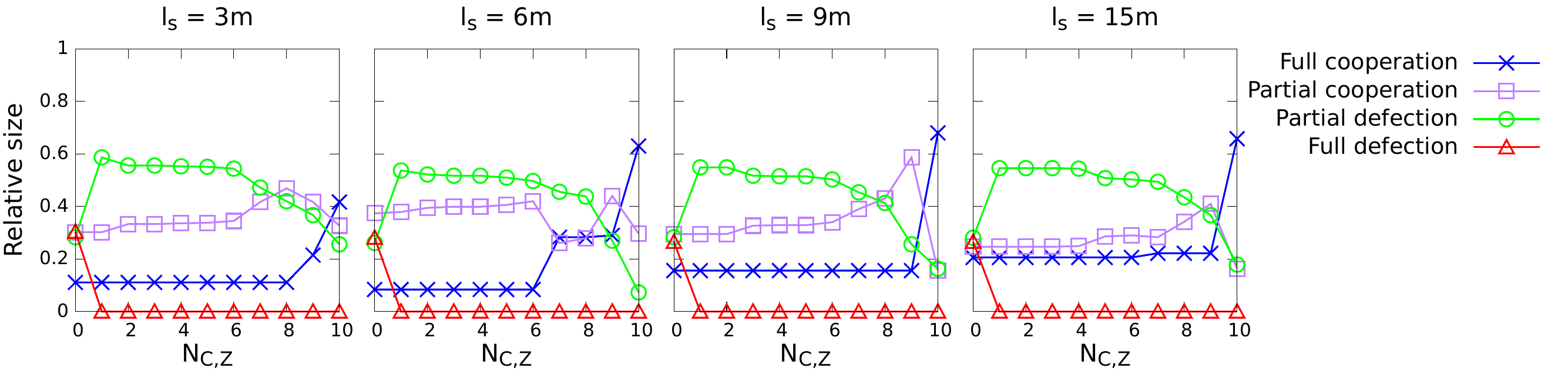}\vspace{-0.3cm}\\
	\end{tabular}	
	\caption{Relative size of different phases in the $(T, S)$ space for different values of sensory range $l_s$: blue cross ($\times$) for the full cooperation phase, purple rectangle ($\square$) for the partial cooperation phase, green circle ($\bigcirc$) for the partial defection, and red triangle ($\bigtriangleup$) for the full defection phase. For different values of $l_s$, the relative size of partial and full defection phases looks qualitatively the same.}
	\label{fig:RelativePhaseSize_ls} 
\end{figure*}

While controlling $\beta$ produces notable difference in the collective helping behavior, change in $l_s$ does not seem to yield any significant changes. Although increasing $l_s$ from 3~m to 6~m improves $\rho_{c, \textnormal{evac}}$ and $P_r$ in the region of SH and PD games when $N_{C,Z}$ is high, further increasing $l_s$ from 6~m does not appear to make considerable improvement. From Figure~\ref{fig:RelativePhaseSize_ls}, it is also observed that the relative size of partial and full defection phases looks qualitatively the same for various values of $l_s$. The relative size of full  cooperation phase is apparently increasing $l_s$ grows while the cooperation phase tends to shrink. It is reasonable to suppose that an injured person might improve the chance of recruiting a volunteer as $l_s$ increase, thus a $(T, S)$ parameter combination in the partial cooperation phase can switch to the full cooperation phase.

\clearpage
\newpage
\section*{References}

\end{document}